\documentclass{emulateapj}
\usepackage{amssymb}
\usepackage{amsmath}
\usepackage{textcomp}
\usepackage{latexsym}

\shorttitle{VLA-COSMOS 1.4~GHz radio population}
\shortauthors{V. Smol\v{c}i\'{c} et al.}

\def\eq#1{\begin{equation} #1 \end{equation}}
\def\comm#1   {{\tt  [COMMENT: #1] }}
\def\f#1   {Fig.~\ref{#1}}
\def\s#1   {Sec.~\ref{#1}}
\def\t#1   {Tab.~\ref{#1}}
\def\ml{multi-wavelength}

\def\mic{$\mathrm{\mu}$m}

\def\sqdeg            {$\Box^{\circ}$}
\def\sqdegs           {$\Box^{\circ}$}
\def\noSN              {2388}

\def\noMULT            {61}
\def\no                {1558} % 1497 single + 61 multi => i_auto < 26 & S/N > 5

\def\noQSO             {139} 
\def\noSF              {340} 
\def\noAGN             {601} 
\def\noGAL             {941}
\def\noUNK             {476} % add multi-comps properly
\def\noNoQSO           {1417} % nogal + nounk
\def\noREM             {830} % 813 single + 17 multi
\def\dn {$\mathrm{D}_\mathrm{n}(4000)$} \def\umag {$u^*$} \def\gmag {$g^+$}
\def\rmag              {$r^+$} % kills the angstrom sign if only \r
\def\imag              {$i^+$} % kills alejo's accent if only \i

\def\uz                {$uz$}
\def\vz                {$vz$}
\def\bz                {$bz$}
\def\yz                {$yz$}
\def\pone              {$P1$}
\def\ponecut           {$0.15$}
\def\ptwo              {$P2$}
\def\na                {$\log{\left( \mathrm{[NII\,\, 6584]/H\alpha}\right)}$}

\def\pp                {(\pone,\ptwo)}
\def\lsun              {$\mathrm{L}_{\odot}$}
\def\RFmethod          {rest-frame color based classification method}
\def\CLASSmethod       {classification method}

\slugcomment{  }

\begin{document}

%\title{ The submillijansky radio population at 20~cm: Separating star forming
%  and AGN galaxies in the VLA-COSMOS survey }

\title{ A new method to separate star forming from AGN galaxies at intermediate
  redshift: The submillijansky radio population in the VLA-COSMOS survey }

\author{V.~Smol\v{c}i\'{c}\altaffilmark{1,2,3},
        E.~Schinnerer\altaffilmark{1}, 
        M.~Scodeggio\altaffilmark{4},
        P.~Franzetti\altaffilmark{4},
        H. Aussel\altaffilmark{5,6},
        M.~Bondi\altaffilmark{7},  
        M.~Brusa\altaffilmark{8},
        C.~L.~Carilli\altaffilmark{9},
        P.~Capak\altaffilmark{3},
        S.~Charlot\altaffilmark{6,10}, 
        P.~Ciliegi\altaffilmark{11},
        O.~Ilbert\altaffilmark{5},  
        \v{Z}.~Ivezi\'{c}\altaffilmark{12},
        K.~Jahnke\altaffilmark{1},
        H.~J.~McCracken\altaffilmark{6}
        M.~Obri\'{c}\altaffilmark{12},
        M.~Salvato\altaffilmark{3},
        D.~B.~Sanders\altaffilmark{5},
        N.~Scoville\altaffilmark{3,13}, 
        J.~R.~Trump\altaffilmark{14},
        C.~Tremonti\altaffilmark{15},
        L. Tasca\altaffilmark{16},
        C.~J.~Walcher\altaffilmark{6},
        G. Zamorani\altaffilmark{11}
        }

\altaffiltext{1}{ Max Planck Institut f\"ur Astronomie, K\"onigstuhl 17,
  Heidelberg, D-69117, Germany } 
\altaffiltext{2}{Fellow of the International Max Planck Research School for
  Astronomy and Cosmic Physics} 
\altaffiltext{3}{ California Institute of Technology, MC 105-24, 1200 East
California Boulevard, Pasadena, CA 91125 }
\altaffiltext{4}{ IASF Milano-INAF, Via Bassini 15, I-20133, Milan, Italy } 
\altaffiltext{5}{ Institute for Astronomy, 2680 Woodlawn Dr., University of
  Hawaii, Honolulu, Hawaii, 96822 } 
\altaffiltext{6}{ Institut d'Astrophysique de Paris, UMR 7095, 98 bis
  Boulevard Arago, 75014 Paris, France }
\altaffiltext{7}{INAF - Istituto di Radioastronomia, via Gobetti 101, 40129
  Bologna, Italy } 
\altaffiltext{8}{ Max Planck Institut f\"ur Extraterrestrische Physik, D-85478
  Garching, Germany }
\altaffiltext{9}{ National Radio Astronomy Observatory, P.O. Box 0, Socorro,
  NM 87801-0387 } 
\altaffiltext{10} { Max-Planck Institut für Astrophysik,
  Karl-Schwarzschild-Strasse 1, 85748 Garching, Germany }
\altaffiltext{11}{INAF - Osservatorio Astronomico di Bologna, via Ranzani 1,
  40127, Bologna, Italy} 
\altaffiltext{12}{ Department of Astronomy,
  University of Washington, Box 351580, Seattle, WA 98195-1580, USA }
\altaffiltext{13}{Visiting Astronomer, Univ. Hawaii, 2680 Woodlawn Dr.,
  Honolulu, HI, 96822} 
\altaffiltext{14}{Steward Observatory, University of Arizona, 933 North Cherry
  Avenue, Tucson, AZ 85721}
 \altaffiltext{15}{Hubble Fellow, University of Arizona,
  Steward Observatory, 933 N. Cherry Ave., Tucson, AZ 85721, USA}
\altaffiltext{16}{Laboratoire d'Astrophysique de Marseille, BP 8, Traverse
du Siphon, 13376 Marseille Cedex 12, France}
\begin{abstract}
  We explore the properties of the submillijansky radio population at
    20~cm by applying a newly developed optical color-based method to separate
    star forming (SF) from AGN galaxies at intermediate redshifts
    ($z\lesssim1.3$).  Although optical rest-frame colors are used, our
    separation method is shown to be efficient, and not biased against dusty
    starburst galaxies.  This \CLASSmethod\ has been calibrated and tested
    on a local radio selected optical sample. Given accurate
    multi-band photometry and redshifts, it carries the potential to be
    generally applicable to any galaxy sample where SF and AGN galaxies are the
    two dominant populations.
  
    In order to quantify the properties of the submillijansky radio
    population, we have analyzed $\sim2,400$ radio sources, detected at 20~cm
    in the VLA-COSMOS survey. $90\%$ of these have submillijansky flux
    densities. We classify the objects into 1) star candidates, 2) quasi
    stellar objects, 3) AGN, 4) SF, and 5) high redshift ($z>1.3$) galaxies.
    We find, for the composition of the submillijansky radio population, that
    {\em SF galaxies are not the dominant population at submillijansky flux
      levels, as  previously often assumed, but that they make up an
      approximately constant fraction of $30-40\%$ in the flux density range
      of $\sim50~\mathrm{\mu}$Jy to $0.7$~mJy}.  In summary, based on the
    entire VLA-COSMOS radio population at 20~cm, we find that the radio
    population at these flux densities is a mixture of roughly $30-40\%$ of SF
    and $50-60\%$ of AGN galaxies, with a minor contribution ($\sim10\%$) of
    QSOs.  
  
\end{abstract}


\section {Introduction}
\label{sec:intro}

The most straight-forward information that can be derived from extra-galactic
radio sky surveys are the radio source counts, which have been extensively
studied in the last three decades \citep{condon84, windhorst85, gruppioni99,
  seymour04, simpson06}. If space was Euclidian, and there was no cosmic
evolution of radio sources, then the differential source counts would follow a
power law with exponent of $2.5$ (see e.g.\ \citealt{peterson97}). Hence, the
observed slope (and the change of the slope) of the radio source counts in
different flux density ranges provides an insight, although quite indirect,
into the global properties of extra-galactic radio sources, and their cosmic
evolution.  Past studies have shown that at 1.4~GHz flux densities above
$\sim100$~mJy the source counts are dominated by 'radio-loud' AGN with
luminosities above the Fanaroff~\& Riley (1974; FR) break
($\sim2\times10^{25}$~W~Hz$^{-1}$; \citealt{willot02}). Decreasing from about
$100$~mJy to $1$~mJy, the source counts follow a power law (e.g.\ 
\citealt{windhorst85}), and are mostly made up of 'radio-loud' objects with
luminosities below the FR break (FR Class~I sources).  However, the
differential source counts change their slope again, i.e.\ they flatten below
1~mJy, and these sub-mJy radio sources have often been interpreted as a rising
new population of objects, which does not contribute significantly at higher
flux densities (e.g.\ \citealt{condon84}).

To date the exact composition of this faint radio population (hereafter
'population mix') is not well determined, and it is rather controversial.
\citet{windhorst85} suggested that the majority of sub-mJy radio sources are
faint blue galaxies, presumably undergoing significant star formation. Optical
spectroscopy, obtained by \citet{benn93}, supported this idea, and the source
counts at faint levels were successfully modeled with a population of
intermediate-redshift star forming galaxies \citep{seymour04}. However,
spectroscopic results by \citet{gruppioni99} suggested that early-type
galaxies were the dominant population at sub-mJy levels. Further, it was
recently suggested and modeled that the flattening of the source counts may be
caused by 'radio-quiet' AGN (radio-quiet quasars and type 2 AGN), rather than
star forming galaxies \citep{jarvis04}; observations support this
interpretation \citep{simpson06}.  Based from the combination of optical and
radio morphology as an identifier for AGN and SF galaxies, \citet{fomalont06}
suggested that at most $40\%$ of the sub-mJy radio sources are comprised of
AGN, while \citet{padovani07} indicated that this fraction
may be $60-80\%$ %only about $20-40\%$ of the faint radio sources are made-up
                 %of star forming galaxies 
(the latter based their SF/AGN classification on a combination of optical
morphologies, X-ray luminosities, and radio--to--optical flux density ratios
of their radio sources).

Two main reasons exist for such discrepant results.  First, the identification
fraction of radio sources with optical counterparts, which is generally taken
to be representative of the full radio population, spans a wide range in
literature (20\% to 90\%) depending on  the depth of both the available
  radio and optical data, as well as the passband used (e.g.\ a larger
  fraction of radio sources are associated with NIR than optical data; see
  \s{sec:CCradio-opt-spitzer} ).  Second, the methods that were used to
separate AGN from SF galaxies have been very heterogeneous in the past,
ranging from pure radio luminosity or morphology cuts, through observed color
properties to optical spectroscopy.

The two main populations of radio sources in deep radio surveys at 1.4~GHz
(20~cm) are active galactic nuclei and star forming galaxies
\citep{condon84, windhorst85}. At this frequency the radio emission
predominantly arises from synchrotron emission powered either by accretion
onto the central super-massive black hole (SMBH) or by supernovae remnants
(e.g.\ \citealt[note that both mechanisms may be at work in a
  given galaxy]{condon92}).  It was shown that radio properties such as the
distributions of mono-chromatic luminosities of SF and AGN galaxies (Seyferts,
LINERs) are comparable and overlapping (at least locally; e.g.\ 
\citealt{sadler99}). Hence, in order to disentangle SF and AGN galaxies in the
radio regime, observations at other wavelengths are required.

Studies of extra-galactic radio sources in the {\em local universe} ($z<0.3$)
have been invigorated due to the recent advent of panchromatic photometric and
spectroscopic all-sky surveys, such as e.g.\ NVSS \citep{condon98}, FIRST
\citep{becker95}, SDSS \citep{york00}, IRAS \citep{beichman85}, 2dF
\citep{colless01} which provide additional panchromatic photometric
(e.g.\ \citealt{simpson06}) and/or optical-IR spectroscopic (e.g.\ 
\citealt{sadler99, best05}) observations.  For example, the panchromatic
properties of radio sources were studied to full detail \citep{ivezic02,
  obric06}, as well as the environmental dependence of radio luminous AGN and
SF galaxies \citep{best04}, and their luminosity function \citep{ sadler99,
  jackson00, chan04, best05}. Further, radio emission as a star formation rate
indicator was well calibrated using a local sample \citep{bell03} and compared
to other star formation tracers \citep{hopkins03}.

However, it still remains to uncover the global properties of the {\em
  intermediate-redshift} ($z\lesssim1.3$) radio sources. For example, the
cosmic star formation history of the universe (i.e.\ the global star formation
rate per unit comoving volume as a function of redshift) was not determined
with a high accuracy using radio data (see e.g.\ \citealt{haarsma00,
  hopkins04}), the radio luminosity function for SF and AGN galaxies at
$z>0.3$ is not known, and the exact composition of the sub-millijansky radio
population is still unknown and a matter of debate \citep{condon84,
  windhorst85, gruppioni99, seymour04, jarvis04, simpson06}. 

In this work and in a number of accompanying papers, we will focus on these
properties of radio sources using the 1.4~GHz VLA-COSMOS survey
\citep{schinnerer07}. The main aim of the current paper is twofold. First, we
develop a method based only on \ml\ photometric data to efficiently separate
SF from AGN galaxies in the VLA-COSMOS 20~cm survey.  Secondly, we use this
classification to derive the composition of the sub-mJy radio population.  In
\s{sec:data} \ we describe the COSMOS \ml\ data, and in \s{sec:cc} \ we
present the cross-correlation of the sources detected at 1.4~GHz with catalogs
at other wavelengths.  In \s{sec:classmet} \ we describe our source
classification methodology and introduce our '\RFmethod' (see below), which we
calibrate and extensively test using a large well-characterized sample of
local galaxies. We present the classification of the VLA-COSMOS 1.4~GHz radio
sources with identified optical counterparts in \s{sec:disentangling} , and in
\s{sec:comparison} \ we compare our \CLASSmethod\ with other classification
schemes proposed in the literature. In \s{sec:resultsRF} \ we study the
'population mix' in the VLA-COSMOS radio survey, based on the entire sample of
VLA-COSMOS radio sources. We summarize our results in \s{sec:summary} .

Throughout the paper we report magnitudes in the AB system, and assume the
following cosmology: $H_0=70,\, \Omega_M=0.3, \Omega_\Lambda = 0.7$.  We
define the radio synchrotron spectrum as $F_{\nu} \varpropto \nu^{-\alpha}$,
and assume $\alpha=0.8$ if not stated otherwise. Hereafter, we refer to our
method to classify the VLA-COSMOS radio sources into five sub-types of objects
(star candidate, QSO, AGN, SF, high-z galaxy) as {\em ``classification
  method''}, and to our method to disentangle only the SF from AGN galaxies,
based on rest-frame color properties, as {\em ``rest-frame color based
  selection method''}.

\section {The multi-wavelength data set}
\label{sec:data}
In this section we describe the COSMOS \ml\ data used for the work presented
here. 
\subsection{ Radio data }
\label{sec:radio_data}

The COSMOS field was observed at $1.4$~GHz ($20$~cm) with the NRAO Very Large
Array (VLA) in A- and C- configuration (VLA-COSMOS Large Project; for details
see \citealt{schinnerer07}).  The final map, encompassing 2\sqdegs, has a
resolution of $1.5'' \times 1.4''$, and a mean $rms$ of $\sim
10.5~[15]$~$\mu$Jy/beam in the central 1~[2]\sqdegs. 

The VLA-COSMOS source catalog reports the peak and total (i.e integrated) flux
density for each object. For extended sources the total flux density is
derived by integrating over the object's size (see \citealt{schinnerer07} for
details), while for unresolved sources it is set to be equal to the peak flux
density.  \citet{bondi07} have shown that bandwidth smearing effects (i.e.\ 
chromatic aberration), combined with the pointing layout of the VLA-COSMOS
observations, systematically decrease the measured source's peak flux density
to $\sim80\%$ of its true value, while the total flux density remains
unaffected. Therefore, to correct for this, all peak flux densities in the
catalog need to be increased by 25\%.  However, such an effect further entails
a necessary re-definition of the sources in the field considered to be
unresolved (cf.\ Fig.~14 in \citealt{schinnerer07} and Fig.~2 in
\citealt{bondi07}).  Therefore, to properly correct for bandwidth smearing
effects, we have re-selected the unresolved sources following \citet{bondi07},
and set their total flux densities to be 1.25 times their peak (respective
integrated) flux densities. Throughout the paper we will use the integrated
flux density, corrected for bandwidth smearing where needed, as the representative
flux density for each source.

In order to minimize the number of possible spurious radio sources ($\sim50\%$
below $5\sigma$), we select only objects from the catalog that were detected
at a signal to noise of $\geqslant5$, and are located outside regions
contaminated by side-lobes from nearby bright sources. This yields \noSN\ (out
of $3643$; i.e.\ $\sim65\%$) sources, 78 of which consist of multiple
components. 

%\comm{Paolo C. gets 2417 sources with $S/N\geqslant5$, which I also get if I
%  don't impose $slob=0$, but I only get 78 (not 80) multi-component sources
%  with $S/N\geqslant5$}

\subsection{ Near-ultraviolet, optical and infrared imaging data } 
The NUV to MIR imaging data and photometry for the COSMOS survey used here
include data taken during 2003--2006 with ground -- (Subaru, KPNO, CTIO, CFHT)
and space (HST, Spitzer) -- based telescopes, covering a wavelength range from
3500~\AA\ to 8~$\mu$m, described them in more detail below.

\subsubsection {Ground-based data}

The data reduction of the COSMOS ground-based observations in 15 photometric
bands ranging from NUV to NIR, and the generation of the photometric catalog,
is presented in \citet{capak07} and \citet{taniguchi07}. Here we make
extensive use of the COSMOS photometric catalog.  The photometric catalog was
selected using the $0.6''$ resolution \imag\ image. However, the photometry
was obtained from the PSF (point-spread function) matched images, which
degrades the resolution to $\sim1.8''$.  The median $5\sigma$ depths in AB
magnitudes in the
catalog for the $u^*$, $B_J$, $g^+$, $V_J$, $r^+$, $i^+$, $i^*$, $z^+$ and
$K_s$ bands\footnote{The '+' super-script and 'J' sub-script designate the
  Subaru filters, while the '*' sign stands for CFHT filters. } are 26.4,
27.3, 27.0, 26.6, 26.8, 26.2, 24, 25.2 and 21.6, respectively (see also tab.~2
in \citealt{capak07}). It is noteworthy that the detection completeness of the
catalog is above $87\%$ for objects brighter than $i=26$.

\subsubsection {Space-based data}

The HST/ACS observations, which covered 1.8\sqdegs\ of the 2\sqdegs\ COSMOS
field, are described in \citet{scoville07} and \citet{koekemoer07}. The F814W
band imaging has a resolution of $0.09''$ and a $5\sigma$ point-source
sensitivity of $I_{AB} = 28.6$ (see also \citealt{capak07}). The ACS source
catalog, which we utilize here, was constructed by \citet{leauthaud07}, with
special care given to the separation of point-sources from extended objects.

The Spitzer observations of the COSMOS field in all seven bands (3.6, 4.5,
5.8, 8.0, 24, 70, 160 $\mu$m) are described in \citet{sanders07}. The 3.6 --
8~$\mu$m band catalog is available to full depth for the entire field.  The
resolution in the 3.6, 4.5, 5.8, 8.0~$\mu$m bands is $1.7''$, $1.7''$,
$1.9''$, and $2''$, respectively. The catalog was generated using SExtractor
on the four IRAC channels in dual mode, with the 3.6~$\mu$m image as the
detection image. The $5\sigma$ depth for point-sources at 3.6~$\mu$m is
1~$\mu$Jy, corresponding to an AB magnitude of 23.9.  In this work we also
make use of the MIPS 24~$\mathrm{\mu}$m catalog obtained from the shallow
observations of the entire COSMOS field during Cycle 2 of the S-COSMOS program
(see \citealt{sanders07} for details). The resolution and $5\sigma$ depth of
the catalog are $6''$ and 0.3~mJy, respectively. The latter corresponds to an
AB magnitude of $17.7$. For the purpose of this paper, we will use only
sources that were detected at 24~$\mathrm{\mu}$m at or above the $3\sigma$
level corresponding to their local rms.

\subsection{ X-ray data }
The full 2\sqdegs\ COSMOS field was observed with the XMM-Newton satellite
EPIC camera for a total net integration time of 1.4~Ms (for a description of
the XMM-COSMOS survey see \citealt{hasinger07}). The limiting flux density of the
XMM-COSMOS survey is $10^{-15}$~erg~cm$^{-2}$~s$^{-1}$ and
$5\times10^{-15}$~erg~cm$^{-2}$~s$^{-1}$ in the soft ($0.5-2$~keV) and hard
($2-10$~keV) bands, respectively. The X-ray point-source detection is
described in \citet{cappelluti07}, and the optical identifications of the
X-ray sources for the first 12 observed XMM fields (over a total of
1.3\sqdegs) are presented by \citet{brusa07}.  For the analysis presented
here we utilize the catalog with 1865 optical counterparts of the XMM-COSMOS
point sources, drawn from the full 2\sqdegs\ XMM-Newton mosaic
\citep{brusa07b}.

\subsection { Photometric redshifts }

The COSMOS photometric redshifts \citep{ilbert07} utilized here are based on
an a large amount of deep multi-color data \citep{capak07b, taniguchi07,
  taniguchi07b}: 6 broad optical bands obtained at the Subaru telescope
($u^+$, $g^+$, $r^+$, $i^+$, $z^+$) and 2 at CFHT ($u^*$ and $i^*$), 8
intermediate and narrow band filters from the Subaru telescope ($IA427$,
$IA464$, $IA505$, $IA574$, $IA709$, $IA827$, $NB711$, $NB816$), deep $Ks$-band
data from the WIRCAM/CFHT camera (McCracken et al., in prep), and $3.6\mu m$
and $4.5 \mu m$ data from the SPITZER IRAC camera \citep{sanders07}. The
photometric redshifts are estimated via a standard $\chi^2$ fitting procedure
\citep{arnouts02} using the code {\it Le
  Phare}\footnote{www.lam.oamp.fr/arnouts/LE\_PHARE.html} written by
S.~Arnouts~\&~O.~Ilbert. A major feature of this method is the calibration of
the photometric redshifts using a spectroscopic sample of $\sim1000$ bright
galaxies ($I_{AB}<22.5$) obtained as part of the zCOSMOS survey
\citep{lilly07}.  We follow exactly the same calibration method as described
in \citet{ilbert06}: a) a calibration of the photometric zero-points, and b)
an optimization of the template SEDs (spectral energy distributions). This
calibration method allows us to remove systematic offsets in the estimates of
the photometric redshifts. A direct comparison between the photometric
redshifts and the zCOSMOS spectroscopic redshifts shows that the photometric
redshifts reach an accuracy of $\sigma\left( \frac{\Delta z}{1+z} \right
)\sim0.014$ at $i<22.5$. The fraction of catastrophic failures is less than
1\% at $i<22.5$. Such an accuracy and robustness can be achieved thanks to
both the intermediate bands and deep NIR photometric data.  The photometric
redshifts for the entire COSMOS population will be described in full detail in
\citet{ilbert07}.  The galaxies in the sample used here are radio selected,
i.e.\ they are not randomly drawn from the global COSMOS population.
Therefore, in \f{fig:zphotspec} \ we show the comparison of the photometric
and spectroscopic redshifts for a sub-sample of our VLA-COSMOS sources with
available spectroscopy (see next section). The accuracy is $\sigma\left(
  \frac{\Delta z}{1+z} \right )=0.027$, which is somewhat lower than
the accuracy for the full sample of COSMOS sources, however it is still
satisfactory.

% 1
\begin{figure}
\includegraphics[bb =  14 14 342 234, width=\columnwidth]{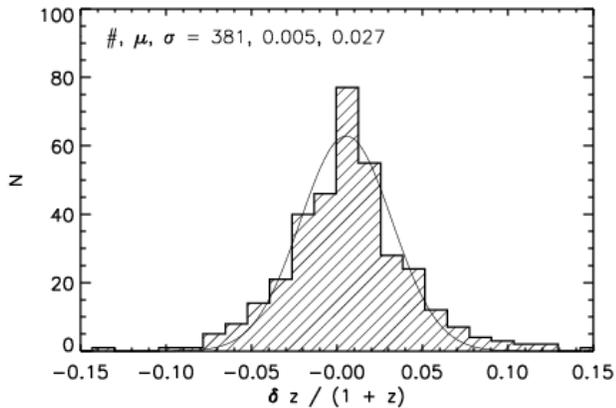} 
\caption{ The comparison of the photometric and spectroscopic redshifts for
  the VLA-COSMOS radio sources with optical counterparts (see
  \s{sec:CCradio-opt} ) for which spectroscopy is available. The distribution
  shown was limited to spectroscopic redshifts $\leq1.3$, consistent with the
  redshift range used in this work. The number of sources ($\#$), mean
  ($\mu$), and standard deviation ($\sigma$) of the distribution are indicated
  in the top left of the panel. Note the excellent accuracy of the photometric
  redshifts.
  \label{fig:zphotspec}}
\end{figure}

As photometric redshift codes generally take into account only galaxy SED
models, the photometric redshifts for broad line AGN are usually poorly
estimated (not better than $\sigma\left( \frac{\Delta z}{1+z} \right
  )\sim0.1$ with a large fraction of catastrophic outliers and no solutions
  found beyond $z\sim1.1$), and alternative ways for their redshift
computations have to be applied. At the time of writing, no accurately
estimated photometric redshifts for broad line AGN exist for the COSMOS
project. The photometric redshifts for broad line AGN will be presented in a
future publication \citep{salvato07}.

\subsection{ Optical spectroscopic data }
\label{sec:optspec}
The ongoing COSMOS optical spectroscopic surveys \citep{trump07,impey07,
  lilly07} provide to date $657$ spectra, with good redshift estimates, for
objects in the VLA-COSMOS 1.4~GHz radio sample described in \s{sec:radio_data}
. We augment this spectroscopic data set with available spectroscopic
information for $65$ galaxies from the SDSS DR4 ``main'' spectroscopic sample,
$13$ objects from the SDSS DR5 quasar catalog \citep{schneider05}, $2$ sources
from the 2dF survey, as well as for $27$ objects taken with the MMT $6.5$~m
telescope, and presented by \citet{prescott06}.  Thus, a total of $764$
spectra is available. However, as a number of sources were spectroscopically
observed multiple times, we have spectroscopic information for $520$ unique
sources in our radio sample.  Throughout the paper, we use the spectroscopic
redshifts, where available. We also use this sub-set of radio sources with
observed optical spectra as a control sample to verify the presented
\CLASSmethod.
%The redshift distribution
%of the available spectroscopic sub-sample is shown in \f{fig:zhisto} \ (light
%grey-shade).  It fairly represents the radio source sample discussed in this
%work up to redshifts of $\sim2$.

%In order to spectroscopically discriminate between star forming galaxies and
%AGN, we utilize diagnostic diagrams, based on emission line flux ratios
%\citep{bpt81,rtt97,kewley01}.  The emission line fluxes were computed by
%Gaussian fits to the emission line profiles (with correspondingly estimated
%errors) after the continuum emission was subtracted. The stellar continuum was
%computed in two different ways. For higher-resolution spectra with excellent
%spectrophotometric calibration (SDSS and COSMOS/VIMOS) stellar synthesis
%models \citep{bc03} were used to describe the stellar continuum emission,
%while for the lower-resolution spectra with poorer spectrophotometric
%calibration (COSMOS/IMACS spectra and spectra presented in
%\citealt{prescott06}) a sliding median filter was used. In \s{sec:BPTs} \ we
%utilize these measurements to test our photometric \CLASSmethod.

\section{ VLA-COSMOS 1.4~GHz radio sources at other wavelengths } 
\label{sec:cc}
In this section we define the {\em 'matched' radio source sample}, a sample of
radio sources with optical counterparts cross-correlated with the panchromatic
COSMOS observations, as well as the {\em 'remaining' radio source sample},
both of which will be used throughout the paper.  First, we restrict the full
VLA-COSMOS radio source sample to objects which have optical counterparts
(Sec.~\ref{sec:CCradio-opt}). Then we positionally match these objects with
sources detected in the MIR (Sec.~\ref{sec:CCradio-opt-spitzer}) and X-ray
(Sec.~\ref{sec:CCradio-opt-X}) spectral ranges. In
Sec.~\ref{sec:CC_no-counterparts} we describe the remaining radio sources that
are either without identified or with identified but flagged optical
counterparts.

\subsection{ Positional matching of the COSMOS radio and NUV/optical/NIR
  catalogs  } 
\label{sec:CCradio-opt}

The VLA-COSMOS Large Project source catalog contains \noSN\ radio sources
detected at a signal to noise $\geqslant5$ and outside side-lobe-contaminated
regions (see \s{sec:radio_data} ). $78$ of these consist of multiple
components.  For the purpose of this paper we match these radio sources with
sources that have also been detected in the optical regime, and are reported
in the COSMOS photometric catalog \citep{capak07}. In order to obtain a sample
with reliable radio-optical counterparts, we positionally match the radio
sources only with optical sources brighter than $i=26$. The reason for this is
illustrated in \f{fig:dist} , where we show the distance between the radio
sources and their nearest optical counterparts as a function of the $i$ band
magnitude. As the median distance rapidly increases for $i>26$ (most probably
introducing a significant number of false match associations) we apply a cut
of $i=26$ to the NUV-NIR photometric catalog before matching the radio and
optical catalogs.

\begin{figure}
\includegraphics[bb =  14 14 352 290, width=\columnwidth]{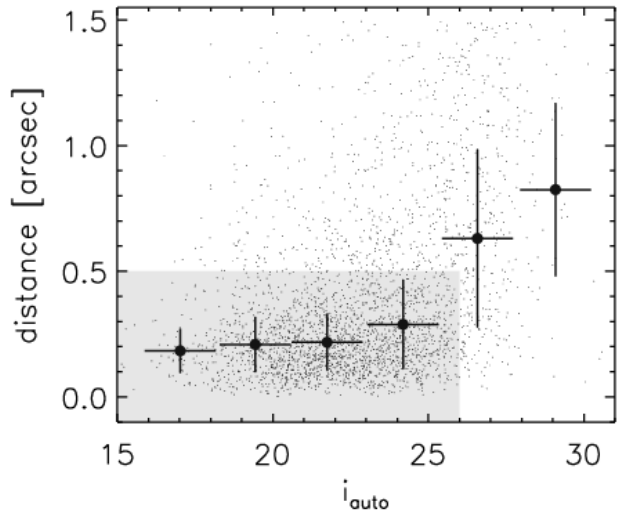} 
\caption{ Distance as a function of $i$ band magnitude for radio sources
  in the VLA-COSMOS survey (with $\mathrm{S/N}\geq5$) positionally matched to
  the closest optical counterpart (small dots). The large  dots show the
  median distance for each magnitude bin (the width of the bin is indicated
  by horizontal lines) and the corresponding interquartile range (vertical
  lines).  Note that the median distance is significantly larger beyond
  $i=26$, presumably introducing a significant number of false match
  associations. The shaded area indicates the allowed matching region, which
  is set by the matching radius ($0.5''$) and the magnitude cut-off ($i=26$)
  used to find secure optical counterparts for the radio sources in the sample
  (see text for details).
  \label{fig:dist}}
\end{figure}

Hence, to find the corresponding optical counterpart for each radio source
(excluding multi-component sources, which are separately addressed below), we
search for the nearest optical neighbor within a radial distance of $0.5''$.
The search radius was chosen in such a way that it balances a high
completeness of true matches and a low false-match contamination rate: A
cut-off of $0.5''$ essentially includes all true matches in the sample, with a
false association rate (computed from the source density in the matched
catalogs) of only $\sim4\%$.
% false association rate = pi * rad**2 * #_optical / area_optical
The high completeness and low contamination are due to the excellent
astrometric accuracy of both the COSMOS radio and optical data.  Our matching
yielded 1749 radio sources with securely identified optical counterparts.
However, 252 ($\sim15\%$) of these are located in masked-out regions (i.e.\ 
around bright saturated stars) in the photometric catalog. Thus, their NUV-NIR
photometry, as well as the photometric redshift computation has a
significantly reduced accuracy. We exclude these objects from our main sample.

The multi-component radio sources in the VLA-COSMOS survey consist of radio
sources which could not be fitted using a single Gaussian function (see
\citealt{schinnerer07}). The radio morphologies of such sources can be fairly
complex (e.g.\ single or double lobed radio galaxies), and this makes it
substantially more difficult to associate such radio sources with the
appropriate optical counterparts (see e.g.\ \citealt{ivezic02, best05}). In
order to avoid any biases which may be caused using an automatic association
procedure, the optical counterparts of the VLA-COSMOS multi-component radio
sources were determined visually. The 1.4~GHz catalog contains 78
multi-component sources detected at or above $5\sigma$, and 65 were securely
associated with an optical counterpart with $i<26$, however 4 are located
inside masked-out areas (around bright saturated stars) in the photometric
catalog, and we therefore exclude them from the main sample.

In summary, the applied matching criteria yield 1814 ($\sim76\%$ of the \noSN\ 
radio sources with $\mathrm{S/N}\ge5$) radio sources with secure optical
counterparts down to $i=26$, 65 of which are multiple component sources. The
most accurate NUV-NIR photometry (i.e.\ excluding flagged regions around
saturated objects) was obtained for \no\ ($\sim 86\%$ of 1814), \noMULT\ of
which are multiple-component radio sources. Hereafter, we refer to this latter
sample of \no\ radio sources, which make up $\sim 65\%$ of the radio sources
with $\mathrm{S/N}\ge5$, and that were matched to the NUV-NIR catalog, as the
{\em 'matched' radio source sample}.  For reference, the 1.4~GHz total flux density
distributions for the complete radio sample, the matched radio sample, and the
sub-sample with available spectroscopy is shown in the top panel in
\f{fig:pophisto} . The distribution of the $i$ band magnitude for the matched
radio sample, as well as the spectroscopic sub-sample, is shown in the bottom
panel in \f{fig:pophisto} . It is also worth noting that our cross-correlation
is consistent with the results of the maximum likelihood ratio technique
applied to VLA-COSMOS sources \citep{ciliegi07a}, however our restrictions for
the masked-out regions in the photometric catalog, as well as the optical
magnitude limit, are more conservative, as the analysis presented here
strongly relies on accurate NUV to NIR photometry.

\begin{figure}
{\center
\includegraphics[bb = 14 14 292 292, width=\columnwidth]{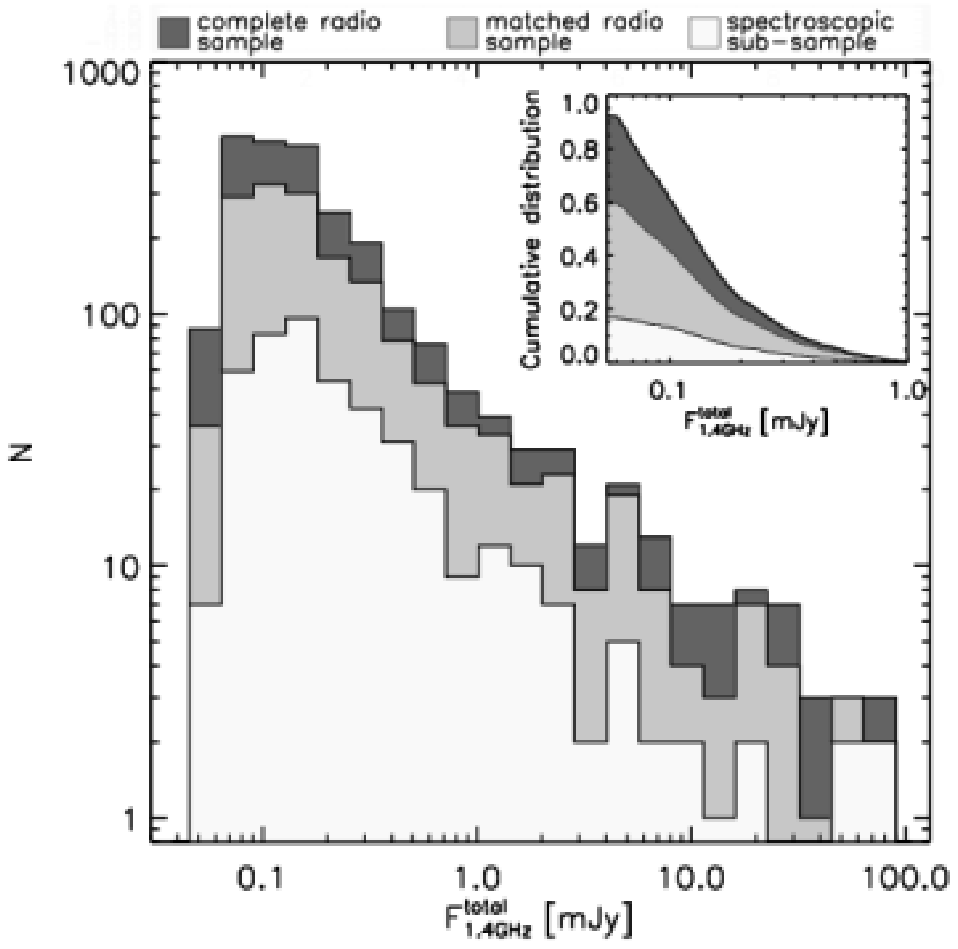} \\
\includegraphics[bb = 14 14 403 292, width=\columnwidth]{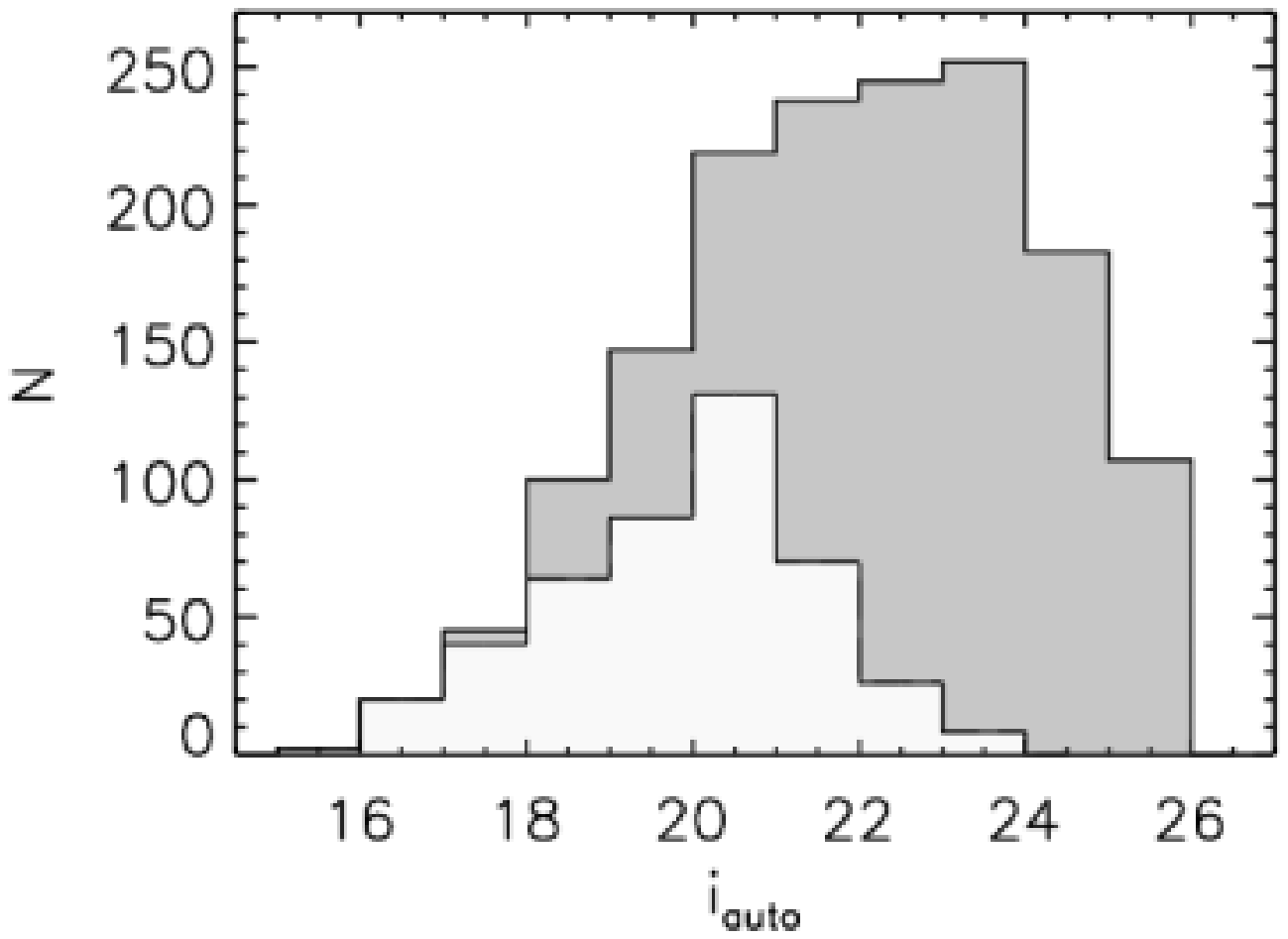} 
\caption{ {\em Top panel}: The distribution of the total 1.4~GHz (20~cm) flux
  density for i) the complete radio source sample ($\mathrm{S/N}\geq5$;
  dark-grey shaded histogram), ii) the matched radio sample (grey-shaded
  histogram) and iii) the spectroscopic sub-sample (light-grey shaded
  histogram). The inset shows the cumulative distribution for the three
  samples computed as a function of {\em decreasing} total flux. Note that the
  spectroscopic sub-sample fairly represents the faint radio population as a
  function of the total flux. {\em Bottom panel}: Distribution of the $i$ band
  magnitude (Subaru where available, otherwise CFHT) for sources in the
  VLA-COSMOS matched radio source sample, and the spectroscopic sub-sample in
  the same notation as in the top panel.
  \label{fig:pophisto}}}
\end{figure}

A further data set that we use in the analysis presented here is the HST/ACS
point-source information.  We extract this information for each radio source
in our matched sample by positionally matching the optical counterparts of the
radio sources with point-sources identified in the HST/ACS F814W source
catalog \citep{leauthaud07}.  Using a matching radius of $0.5''$
yields 47 objects in our matched radio sample classified as point sources
based on the HST/ACS F814W images. The mean distance
between the matched objects is only $(0.12\pm 0.07)''$.

\subsection{ Radio -- optical sources with IRAC and MIPS detections }  
\label{sec:CCradio-opt-spitzer}

We cross-correlate the matched radio source sample with the S-COSMOS -- IRAC
catalog using a maximum allowed distance to the optical counterparts of our
radio sources of $0.5''$. [Note that such a cross-correlation allows for a
maximum distance between the radio and MIR sources to be $1''$.] Such an
adopted search radius essentially selects a complete radio -- optical -- MIR
sample with a false match association for the MIR sources of $\lesssim1$\% with
the optical counterparts, and $\lesssim4$\% with the radio counterparts. In
summary, out of \no\ radio sources in the matched radio sample, $1448$
($93\%$) have secure MIR counterparts.

The 24~\mic\ flux densities for all our radio sources were obtained from the COSMOS
field observations using the S-COSMOS -- MIPS shallow survey with a resolution
of $6''$. Although a relaxed search radius of $5''$ was used to find the radio
-- 24~\mic\ counterparts, the median distance is only $0.19''$ with an
interquartile range of $0.16''$. About $50\%$ (799 out of 1558) sources in the
matched radio sample have a MIPS counterpart at 24~\mic\ with a signal to
noise at or above 3.

\subsection{ Radio -- optical sources with point-like X-ray emission } 
\label{sec:CCradio-opt-X}
Using the maximum likelihood ratio technique \citet{brusa07} presented the
optical identifications of the X-ray point-sources \citep{cappelluti07}
detected in the XMM-COSMOS survey \citep{hasinger07}. Here we utilize their
identifications to match the sources in our matched radio sample with detected
X-ray point sources. Out of \no\ radio sources with optical counterparts, 179
($12\%$ of 1558) are identified as point-sources in the X-ray bands. 17 of
these
% flag = 10
have multiple counterpart candidates as defined by \citet{brusa07}.  In these
cases, if we assume that the radio sources are physically associated with the
X-ray sources, then the radio data, which have a significantly better
astrometric accuracy, can be used to constrain more precisely the optical
counterpart of this given object. A visual inspection of the 17 sources,
classified as having ambiguous identifications by \citet{brusa07b}, strongly
suggests that their most probable optical counterparts, reported in the X-ray
-- optical catalog, are real associations.  Hence, we proceed in our analysis
taking all 179 X-ray detected point sources to be true counterparts of the
objects in the matched radio sample.

\subsection{ Radio sources with photometrically flagged or without optical
  counterparts at other wavelengths }
\label{sec:CC_no-counterparts}

In \s{sec:CCradio-opt} \ we have defined the matched radio sample which
consists of \no\ 1.4~GHz sources that have optical counterparts out to an $i$
band magnitude of 26, and within a radial distance of less than $0.5''$. These
sources were also required to have the most accurate NUV-NIR photometry, i.e.\ 
counterparts within flagged regions due to saturation and blending effects in
the NUV-NIR images were excluded. Thus, \noREM\ radio sources remain with no
identified optical counterparts within these limits, 256 (i.e.\ $\sim30\%$) of
which have counterparts with $i\leq26$ that lie in masked-out regions.
Hereafter, we will refer to this sample of sources as the {\em 'remaining
  radio source sample'}.  We positionally match these sources to the IRAC
catalog using a maximum allowed distance of $1''$, and find 610 ($\sim75\%$)
matches. Based on Poisson statistics and the source density of the MIR
sources, such a search radius essentially includes all true matches with a
false contamination rate of $\lesssim4$\%. It is worth noting that more than
one half of the remaining $25\%$ of the radio objects were independently
identified as possible spurious sources, based on visual inspection, while the
other half are either located in blended regions in the IRAC images or
slightly further away than the allowed $1''$ from the position reported in the
IRAC catalog (the morphology of the IRAC sources being often extended). Thus,
we consider these $\sim75\%$ of radio -- MIR matches representative of the
entire remaining radio population. Out of the \noREM\ sources 318 (i.e.\ 
$\sim40\%$) have MIPS 24~\mic\ detections ($\mathrm{S/N}\geq3$), and 31
($\sim4\%$) have XMM point source counterparts (these 31 objects are a
sub-sample of the 256 objects in the flagged regions).  In \s{sec:resultsRF} \ 
we analyze the properties of these remaining sources, and their contribution
to the 'population mix' in the VLA-COSMOS survey. The summary of the \ml\ 
cross-correlation of the VLA-COSMOS radio sources is given in \t{tab:class} .

\section{ Classification methodology }
\label{sec:classmet}

Extragalactic radio sources consist of two main populations: star forming and
AGN galaxies. We further divide the AGN class into two  sub-classes:
QSOs (often unresolved in optical images, with broad emission lines in their
spectra and high optical luminosity) and objects where the AGN does not
dominate the entire SED, such as Type 2 QSOs, low-luminosity AGN (Seyfert and
LINER galaxies) and absorption-line AGN (resolved in optical images, with both
broad, narrow or no emission lines in their optical spectra).  Throughout the
paper, we will mostly refer to the latter sub-class only as 'AGN'.
%
%
%\subsection{ The locally tested method  }
%\label{sec:locmet}
%
%In this section we calibrate our \RFmethod\ using a large sample of well known
%local galaxies with available high-resolution ($\mathrm{R}=1800$) optical
%spectroscopy, drawn from the SDSS, NVSS and IRAS sky surveys.  In
%\s{sec:RFoutline} \ we present the background and motivation for the
%\RFmethod. In \s{sec:p1p2} \ we describe the rest-frame colors used for this
%method, and in \s{sec:locsample} \ we introduce the local sample.  The
%completeness and contamination of the selected samples of SF and AGN galaxies
%is derived in \s{sec:loccompl} , and in \s{sec:ulirgbias} \ we test whether
%our \RFmethod\ is biased against dusty starburst galaxies.
%
%\subsubsection{Outline}
%\label{sec:RFoutline}
%
%Spectroscopic diagnostic diagrams are the optimal tools for separating SF and
%AGN galaxies (i.e.\ predominantly Seyfert and LINERs), and they have already
%been proposed by \citet{bpt81}.  

The commonly adopted and well-calibrated tool for disentangling SF galaxies
from low-luminosity AGN (Seyfert and LINERs) is the optical spectroscopic
diagnostic diagram \citep{bpt81}, that is based on two emission line flux
ratios ([OIII 5007]/H$\beta$ vs.\ [NII 6584]/H$\alpha$; hereafter {\em BPT
  diagram}; see also \citealt{veilleux87, rtt97, kewley01}).  This diagnostic
tool has been extensively used in the past for a successful separation of
local SF and AGN galaxies \citep{sadler99, kauffmann03, brinchmann04, obric06,
  smo06}.  However, as spectroscopic observations are very expensive in terms
of telescope time, especially when large numbers of faint objects need to be
observed, alternative methods for the separation of SF from AGN galaxies, that
eliminate the need for spectroscopy, have to be invoked. Here we develop such
a method (hereafter '\RFmethod'), which we apply in the next sections to
identify SF and AGN galaxies in the VLA-COSMOS matched radio sample.
The main idea of our method is drawn from the findings that the overall
NUV to NIR SED of galaxies is a one-parameter family, and that
spectral diagnostic parameters, such as line strengths, appear to be well
correlated with the overall galaxy's SED (see \citealt{obric06,smo06}). In
particular, \citet{smo06} have found {\em a tight correlation between
  rest-frame colors of emission-line galaxies and their position in the BPT
  diagram}.  This correlation thus provides a powerful tool for disentangling
SF from AGN galaxies using only photometric data, i.e.\ rest-frame colors, and
we utilize it as the key of our \RFmethod.

\subsection{ Calibration of the \RFmethod\ in the local universe }
\label{sec:locmet}

%\subsubsection{The rest-frame color}
%\label{sec:p1p2}

%\citet{smo06} used the SDSS main galaxy sample to synthesize rest-frame
%magnitudes in the modified Str\"{o}mgren photometric system (\uz,\vz,\bz,\yz\ 
%encompassing the wavelength range of 3200 -- 5800\AA;
%\citealt{odell02})\footnote{The filter response curves of the modified
%  Str\"{o}mgren system are available at http://www.mpia-hd.mpg.de/COSMOS/},
%and study the rest-frame color properties of these galaxies. In order to
%optimally quantify the distribution of galaxies in the rest-frame color-color
%space, they defined a set of principal component axes (\pone,\ptwo), where
%\pone\ measures the position along the galaxy locus, and \ptwo\ the position
%perpendicular to it (see Fig.~4 in \citealt{smo06}). A more detailed
%description of the colors, including their equational form, is given in
%Appendix~\ref{app:p1p2}. Here we utilize only the \pone\ color that strongly
%correlates with emission line properties of AGN and SF galaxies (see
%\s{sec:loccompl} ).

\subsubsection{ The local spectroscopic sample }
\label{sec:locsample}
In order to obtain insight into the efficiency of the classification method,
based on the rest-frame color (\pone ; see below), we construct a large
sample of well-known low-redshift galaxies ($0.01<z<0.3$), whose properties
are assumed to present well the properties of the galaxies in the VLA-COSMOS
matched radio sample.  The local sample was generated from the SDSS main
galaxy sample, positionally matched to sources detected in the 1.4~GHz NVSS
survey.  Additionally, a sub-sample of galaxies detected with IRAS was
constructed (see \citealt{obric06} for details about the cross-correlation of
the SDSS, NVSS, IRAS catalogs). The SDSS/NVSS sample contains 6966 galaxies
and the SDSS/NVSS/IRAS sample 875 galaxies with available SDSS optical
spectroscopy.  The computation of rest-frame colors (\pone,\ptwo) for these
galaxies is presented in \citet{smo06}. The rest-frame colors \pone\ and
\ptwo\ optimally quantify the distribution of galaxies in the rest-frame
color-color space. They are derived from the modified Str\"{o}mgren
photometric system (\uz,\vz,\bz,\yz\ encompassing the wavelength range of 3200
-- 5800\AA; \citealt[see also \citealt{smo06}]{odell02}) via a 2-dimensional
principle component analysis in color-color space. A more detailed description
of the colors, including their equational form, is given in
Appendix~\ref{app:p1p2}.

It is noteworthy to mention that given a) the
detection limits, and b) the areal coverage of the NVSS and VLA-COSMOS
surveys, both surveys observe approximately the same populations of objects,
although over different redshift intervals (see Fig.~1 in
\citealt{schinnerer07}). Assuming that evolutionary effects with redshift do
not significantly alter the reliability of the identification method presented
below, at least out to $z=1.3$, this makes the local sample of galaxies
representative of the galaxies in the VLA-COSMOS matched radio sample.

Based on {\em spectral line} properties we separate the local sample into
three classes of objects: {\em AGN, star-forming galaxies and composite
  objects}, where the latter are considered to have a comparable contribution
of both star formation and AGN activity. First, galaxies {\em with emission
  lines in their spectra} are separated into these three classes using their
position in the BPT diagram (see bottom panel in \f{fig:bptsdssnvss}
).\footnote{Note that the classification of AGN, SF, and composite galaxies
  based on the BPT diagram changed compared to the one used in \citet{smo06}.
}  Second, as the galaxies that have
{\em no emission lines} in their spectra cannot be star forming (see also
\s{sec:ulirgbias} \ for a discussion of this point), and as all of the objects
in the SDSS/NVSS sample are observed to have 1.4~GHz emission which arises
either from AGN or star formation activity in a galaxy, we define galaxies
without emission lines in their spectra as AGN (see also \citealt{best05} who
classified these types of objects as absorption line AGN).

% 4
\begin{figure}
{\center
\includegraphics[bb = 14 14 265 402, width=\columnwidth]{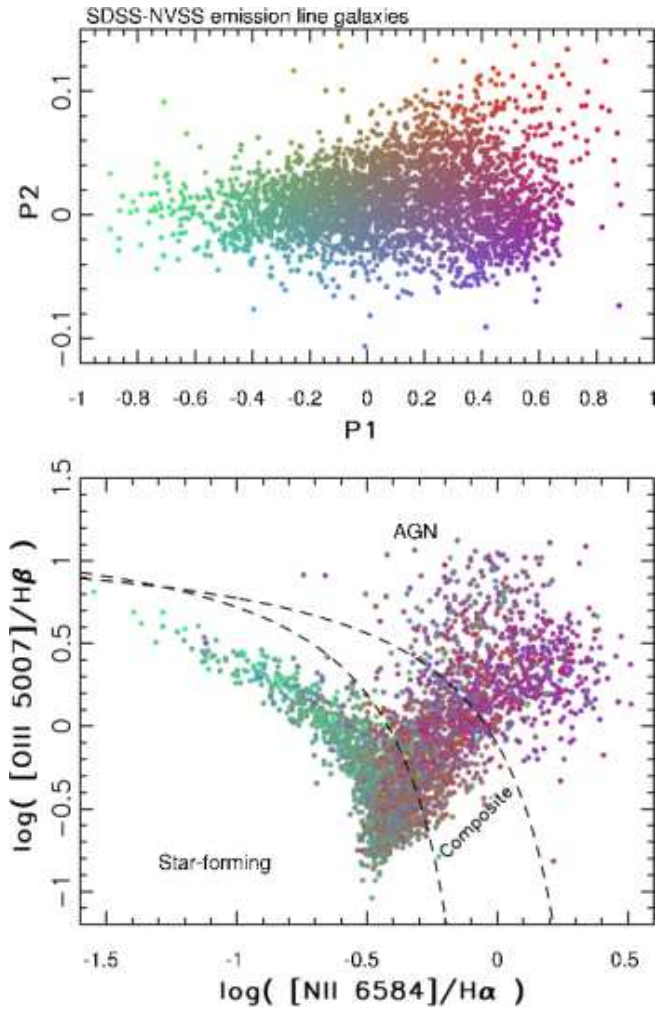}
\caption{ {\em Top panel:} The distribution of $\sim3,400$ SDSS emission
  line galaxies, in the redshift range of $0.01$ to $0.3$, drawn from the DR1
  ``main'' spectroscopic sample, which were also detected by the 1.4~GHz NVSS
  survey, in a diagram constructed with the principal rest-frame colors
  (\pone,\ptwo).  Each dot corresponds to one galaxy, and the color code is
  determined by the position in this plane.  The {\em bottom panel} shows the
  distribution of these galaxies in the (BPT) diagram \citep{bpt81}. The lower
  dashed line separates the regions populated by star-forming and composite
  galaxies \citep{kauffmann03}. The upper dashed line separates the regions
  populated by composite and AGN galaxies \citep{kewley01}.  The dots are
  colored according to their \pone\ and \ptwo\ values given in the top panel.
  Note the strong correlation between the rest-frame colors and the emission
  line flux ratios (see text for details).\label{fig:bptsdssnvss}}}
\end{figure}

\subsubsection{ Completeness and contamination due to the photometric
  selection} 
\label{sec:loccompl}

In the top panel in \f{fig:bptsdssnvss} \ we show the distribution of
$\sim3,400$ SDSS/NVSS emission line galaxies in the \pp\ rest-frame
color-color diagram.  The color code is determined by the position of a galaxy
in this plane. The bottom panel shows the BPT diagram for the same galaxies
with the colors adopted from the upper panel. For these radio luminous
galaxies, a strong correlation exists between their rest-frame optical colors
and emission line properties, in particular between \pone\ and \na. We want to
stress that the SDSS/NVSS galaxies with no emission lines in their spectra
have typically red \pone\ colors, with a median \pone\ value of 0.46 and an
interquartile range of 0.13 (see also e.g.\ Fig.\ 12 in \citealt{smo06}). This
implies that the rest-frame color \pone\ can be used as an efficient separator
between AGN and SF galaxies in samples where these two types of galaxies are
the two dominant populations.

\citet{smo06} argued that the \ptwo\ color is a proxy for the galaxies' dust
content, with higher values of \ptwo\ corresponding to higher dust
attenuation.  However, as the dynamic range of \ptwo\ is very narrow, we will
rather base our selection on the \pone\ color alone. Hereafter, based on a
compromise between completeness and contamination of the photometrically
selected samples of galaxies, we choose a color-cut of \pone$=$\ponecut\ as
the dividing photometric value between star forming galaxies
(\pone$\leq$\ponecut ) and AGN (\pone$>$\ponecut).  Given this boundary we
infer that the sample of photometrically selected star forming galaxies is
$85\%$ complete, and contaminated by AGN and composite objects at the $20\%$
and $10\%$ levels, respectively. Vice versa, the photometrically selected
sample of AGN is $90\%$ complete and contaminated by SF and composite galaxies
at the $5\%$ and $15\%$ level, respectively (see Appendix~\ref{app:complfull}
for computational details).  We will use these estimates in the following
analysis to statistically correct the photometrically selected SF and AGN
samples in the VLA-COSMOS survey.

\subsubsection{ Is the rest-frame color based selection method biased against
  dusty starburst galaxies?}
\label{sec:ulirgbias}

One of the main advantages of radio observations is that the intrinsic
physical properties that drive the radio emission can be derived without any
need for dust-extinction corrections (as radio emission passes freely through
dust). In particular, radio observations provide a dust-unbiased view of
star-formation (see \citealt{condon92} for a review; see also
\citealt{haarsma00}).  Hence, it is important to address whether our
rest-frame color selection technique misses out dusty starburst galaxies. In
order to do this we study a sub-sample of SDSS/NVSS galaxies that were also
detected with the IRAS satellite at IR wavelengths.

First we address the composition of the 'missed' $\sim15\%$ of the SF galaxies
(note that the photometrically selected SF galaxy sample is $85\%$ complete)
in order to show that our selection does not introduce biases against the most
luminous starburst galaxies.  About $30\%$ of the SF galaxies with
\pone$>$\ponecut\ have IRAS detections, which is consistent with the fraction
of IRAS detections in the SF galaxy sample with \pone$\leq$\ponecut. This
suggests that the composition of the {\em missed} SF galaxies is not
significantly different from the composition of the {\em selected} SF
galaxies.  Second, the fractions of luminous and ultra-luminous IR galaxies
(LIRGs and ULIRGs) in the spectroscopically classified SF galaxy samples with
\pone$\leq$\ponecut\ and \pone$>$\ponecut\ 
appear to be consistent with each other. %Thus, the \RFmethod\ does not
%introduce a bias against dusty starburst galaxies.
And third, the completeness and contamination of the SDSS/NVSS/IRAS galaxy
sub-samples selected using the \RFmethod\ is fairly consistent with the
properties of the entire SDSS/NVSS sample (see Appendix~\ref{app:compl} for
details). All of this implies that the selection criteria for SF and AGN
galaxies adopted on the basis of the analysis of the full SDSS/NVSS sample
works almost equally efficiently for the IRAS detected sub-sample. Even
further, only $\sim10\%$ and $\sim5\%$ of star forming LIRGs and ULIRGs
galaxies, respectively, are omitted by the method (see
Appendix~\ref{app:compl} for details).

It is noteworthy that in the entire SDSS/NVSS/IRAS sample only 48 objects
(i.e.\ $\sim5\%$) were identified as absorption line AGN, i.e.\ having no
emission lines in their optical spectra.  It is possible that very high dust
obscuration may suppress the detection of emission lines in the optical
spectrum.  A visual search for signatures of HII regions in the
48 SDSS color-composite images suggested that at the most $30\%$ of these
galaxies may possibly be undergoing star formation (e.g.\ possible galaxy
merger, or extended morphology). Therefore, only a negligible fraction of less
than $1.5\%$ of the SDSS/NVSS/IRAS galaxies may be so heavily dust-obscured
that no emission lines would be detected in their optical spectra.

In order to test this issue further, we have synthesized the \pone\ color for
the 'standard' dusty starburst galaxies M~82 (a typical LIRG), and Arp~220
(the prototypical ULIRG) using spectral templates given by Polletta et al.\ 
(2006). The optical to NIR part of these templates was generated using the
stellar population synthesis code GRASIL \citep{silva98}.  The derived \pone\ 
colors for M~82 and Arp~220 are 0.078 and 0.149, respectively, implying that
M~82 would not have been missed by the \RFmethod, while Arp~220, although
close to the adopted limit in \pone, still lies within our selection criterion
(note that in reality observational photometric errors of M~82- and Arp~220-
like objects will introduce a scatter in P1; see also \s{sec:testRF} ).  Based
on the tests presented above we conclude that our \RFmethod\ is not
significantly biased against dusty starburst galaxies.

\subsection{ Application of the \RFmethod\ to the multi-wavelength
  photometry of the VLA-COSMOS radio -- optical galaxies }
\label{sec:synthesis}

%In \s{sec:locmet} \ we have presented our \RFmethod, which we have calibrated
%using a local galaxy sample, drawn from the SDSS and NVSS sky surveys. Using
%galaxies from this sample, that were also detected by the IRAS satellite in
%the IR, we have shown that the method is not significantly biased against
%selecting dusty starburst galaxies (star forming LIRGs and ULIRGs). This is an
%important point for a radio sample as radio emission is an efficient tracer of
%star formation, and of particular importance especially in the most dusty
%systems. Therefore,

In the previous section we have  studied, and  calibrated, the
\RFmethod\ using the local galaxy sample.  Assuming that evolutionary effects
with redshift do not significantly affect the reliability of the
classification, we can safely apply it to the galaxies in the VLA-COSMOS
1.4~GHz matched radio sample (\s{sec:RF} ) \  to separate SF from AGN
galaxies at intermediate redshifts. However, first we need to derive the
rest-frame \pone\ color from the observed SED of the VLA-COSMOS galaxies. We
do this via high-resolution SED fitting, described in \s{sec:p1derivation} ,
and we test the accuracy of the rest-frame color synthesis in \s{sec:testRF} .

\subsubsection { Derivation of rest-frame colors } 
\label{sec:p1derivation}

In order to estimate the rest-frame color \pone\ for each galaxy in the
matched radio sample that we do not classify as a star or QSO (see
\s{sec:disentangling} ), we use the GOSSIP (Galaxy Observed Simulated SED
Interactive Program) software package \citep{franzetti05}, designed for
fitting a galaxy's SED to a set of chosen spectral models.  The SED of the
galaxies in our sample, that we use for fitting, extends from $3500$~\AA\ to
$2.5~\mathrm{\mu}$m (comprised in 6 photometric pass-bands) and we fit
to each observed SED a realization of $\sim100,000$ spectra built using the
\citet{bc03} stellar synthesis evolutionary models.
%The library of model spectra was parameterized similarly
%to \citet[see also \citealt{kauffmann03}; \citealt{kong04}]{salim07}.
Star formation histories have been parameterized by an underlying continuous
star formation history (decaying exponentially), and randomly superimposed
bursts (see also \citealt{kauffmann03, kong04, salim07}). We cover ages
between $100$~Myr and $13.5$~Gyr, specific star formation rates
(star formation rate per unit galaxy stellar mass) between $10^{-15}$
yr$^{-1}$ and $3.93\times10^{-8}$ yr$^{-1}$ and metallicities from a tenth to
twice solar.
%The approach of fitting the multi-wavelength broad-band SED to obtain physical
%parameters has been validated at low redshifts in Salim et al.  (2007).

For each object in our sample the model spectra in our library are redshifted
to the galaxy's measured redshift (spectroscopic where available, otherwise
photometric), then each spectrum is convolved with the observed filter
response function\footnote{The COSMOS filter response curves can be found
  here: http://www.astro.caltech.edu/$^\sim$capak/cosmos/filters}, and then
fitted to the available observed photometric data, using a direct $\chi^2$
minimization procedure.  Output parameters, such as e.g.\ rest-frame colors,
stellar mass or the 4000~\AA\ break, are taken from the best fit model
spectrum. In order to derive physically meaningful output parameters, we
restrict the fitting procedure to models that have an age smaller than that of
the Universe at the galaxy's redshift.

%\subsubsection{ Tests on the derived rest-frame colors }
\subsubsection{ Accuracy of the derived rest-frame colors }
\label{sec:testRF}
%In this section we test the accuracy of the synthetic magnitudes and colors
%derived via SED fitting (see previous section) in order to assess the expected
%uncertainties of the \pone\ color synthesis for the galaxies in the VLA-COSMOS
%matched radio sample.
%
%The accuracy of the synthesized (relative to observed) magnitudes and colors
%is summarized in \f{fig:magcol} \ for {\em all} galaxies in the matched radio
%sample (objects that were not classified as stars or QSOs; see
%\s{sec:disentangling} ).  The colors and magnitudes are reproduced with a
%satisfying accuracy ($\sim0.07$ and $\sim0.04$, respectively). The synthesized
%$V_J$ shows a slight systematic offset ($\sim0.05$), which may arise from the
%incomplete knowledge of the filter response curve and/or the presence of
%strong spectral emission lines which are not taken into account in the model
%spectra.  Nonetheless, we can conservatively conclude that the overall
%accuracy of the synthetic magnitudes and colors is $\sim0.1$.  It is also
%noteworthy that \pone\ is derived from a rest-frame spectral range of
%$3800-5800\AA$, which corresponds to observed $V$, $r$, $i$ and $z$ band
%ranges for redshifts of roughly $0.1$ to $0.9$.  Thus the obtained color and
%magnitude accuracies in these observed spectral ranges mimic adequately the
%rest-frame spectral range of interest.

The synthesized, relative to observed, colors and magnitudes for the
VLA-COSMOS galaxies, computed as described above, are reproduced with a
satisfying accuracy of $\sim0.07$ and $\sim0.04$, respectively.  To further
test the accuracy of the derived rest-frame color, we synthesize P1 for a
sample of $\sim1700$ local SDSS/NVSS galaxies. For these galaxies the P1
colors, computed from their spectrum (with an accuracy of better than 0.03
magnitudes; see \citealt{smo06}), are also available. Thus, comparing \pone\ 
synthesized via SED fitting with the reference \pone\ derived from the
spectrum gives us a direct measure of the achieved accuracy, shown in
\f{fig:p1sdssgossip} \ (top panel). The root-mean-square-scatter is $\sim0.1$.

\begin{figure}
{\center
  \includegraphics[bb = 14 14 292 357, width=\columnwidth]  {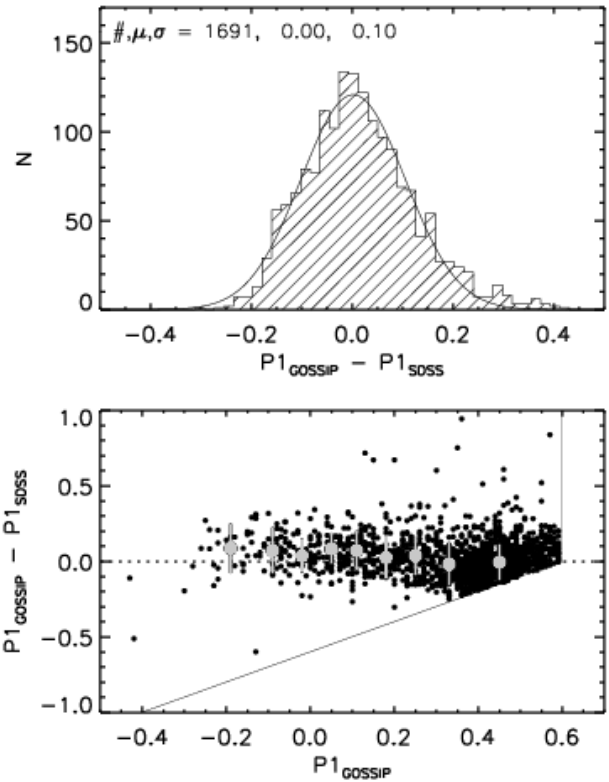}
 \caption{  The comparison of the \pone\ color synthesized from the SDSS
   spectrum ($\mathrm{P1_{SDSS}}$ that has a typical error of 0.03) and via
   SED fitting ($\mathrm{P1_{GOSSIP}}$) for $\sim1700$ SDSS/NVSS galaxies from
   the DR1 ``main'' spectroscopic sample (top panel). The \pone\ color is
   reproduced with a spread of $\sim0.1$. The bottom panel shows the
   difference in the \pone\ color as a function of $\mathrm{P1_{GOSSIP}}$. The
   thin solid lines show the upper \pone\ limits of 0.6, imposed for this
   analysis. Note the slight systematic trend in the derived \pone\ color as a
   function of $\mathrm{P1_{GOSSIP}}$ (the large dots are the median offsets,
   with indicated $1\sigma$ error bars).  The trend presumably arises due to
   the presence of strong emission-lines in the rest-frame wavelength range of
   $3800 - 5800$~\AA, that are not taken into account in the BC03 models. We
   use these offsets to correct for these systematics in the derivation of
   \pone\ for the clean radio sample (see text for details).
   \label{fig:p1sdssgossip}}}
\end{figure}

In the bottom panel in \f{fig:p1sdssgossip} \ we show the difference of the
\pone\ color as a function of the synthesized (GOSSIP) \pone\ color for the
SDSS/NVSS galaxies. A slight systematic trend is present as a function of the
derived \pone\ color. This trend presumably arises from the presence of strong
emission lines in the galaxies' SEDs (in the rest-frame wavelength range that
is used to derive \pone ), which are not taken into account in the BC03 model
spectra.  In our further analysis of the VLA-COSMOS data, we use the median
offset, shown in the bottom panel in \f{fig:p1sdssgossip} , to correct the
derived \pone\ color, and we consider the synthetic \pone\ color to be
accurate to $\sim0.1$~mag. However, an error of 0.1~mag for 68\% of the
galaxies, and 0.2 for 95\%, may substantially alter the SF/AGN selection,
introducing the largest uncertainty for the galaxies that have \pone\ colors
close to the chosen \pone\ boundary.  To account for these uncertainties, we
simulate the \pone\ errors using a randomly drawn Gaussian distribution with a
standard deviation of 0.1 centered at zero (see top panel in
\f{fig:p1sdssgossip} ). These errors are then added to the galaxies' \pone\ 
colors derived from the best fit template in the SED fitting, and the SF/AGN
selection (see \s{sec:RF} ) is applied. By repeating this procedure 10,000
times we obtain a robust statistical distribution of the number of selected SF
and AGN galaxies yielding $356$ SF and $585$ AGN galaxies (mean)
with a root-mean-square scatter of only $7$. %We reach the same result if we
%model the error distribution using two separate Gaussian distributions for
%blue and red galaxies.

Applying our SF/AGN selection using the P1 distribution obtained from the best
fit template in the SED fitting yields \noSF\ SF and \noAGN\ AGN galaxies (see
\s{sec:RF} ).  Thus, $\sim5\%$ less SF galaxies are selected.  This is easily
understood as the blue tail of the \pone\ distribution contains a smaller
number of galaxies than there are in the prominent red peak (see e.g.\ 
\f{fig:p1sdssnvss} ). Therefore, a normal error distribution will
systematically scatter more galaxies to the blue \pone\ region, than to the
red one. We conclude that the photometric errors of the synthesized \pone\ 
color introduce a number uncertainty of $\sim5\%$ in favor of SF galaxies.
Although $\sim5\%$ is not significant, it is necessary to keep this bias in
mind in the analysis of the 'population mix' of submillijansky radio sources
(\s{sec:popmix} ).

\section{ Classification of VLA-COSMOS sources in the matched radio
   sample }
\label{sec:disentangling}

In this section we present the classification of the entire VLA-COSMOS matched
radio sample into star candidates (\s{sec:stars} ), QSOs (\s{sec:qso} ), SF,
AGN and high-z galaxies (\s{sec:RF} ). We begin with a summary of the
sample definitions (\s{sec:nom} ).

\subsection{ Outline and nomenclature }
 \label{sec:nom}
 
 In \s{sec:locmet} \ (see also Appendix~\ref{app:complmain}) we have presented
 the calibration and effectiveness of the \RFmethod\ for separating {\em
   galaxies} dominated by star formation from those dominated by AGN activity.
 For this we have used the SDSS ``main'' spectroscopic sample -- a pure {\em
   galaxy} sample that by definition does not contain any star-like objects
 \citep{strauss02}. This, obviously, implies that the same effectiveness of
 the method can only be reached if the \RFmethod\ is applied to a comparable
 sample, i.e.\ {\em galaxies only}.  However, our VLA-COSMOS matched radio
 sample consists not only of galaxies, but also of stellar like sources, where
 the latter are either stars or quasi stellar objects (QSOs). Therefore, we
 classify the sources in the matched radio sample into five sub-types -- a)
 star candidates, b) quasi stellar objects (QSOs), c) active galactic nuclei
 (AGN), d) star forming (SF), and e) high redshift ($z>1.3$; high-z) galaxies.
 The latter three sub-types compose our {\em ``VLA-COSMOS galaxy sample''}.
 The properties of each sub-type are summarized as follows.
\begin{itemize}
\item[] {\bf Stars:} Point-sources in the optical, with their SEDs best fit
  using a stellar template.
\item[] {\bf OSQs:} Point-sources in the optical (stellar-like SEDs are
  excluded; see above). This criterion essentially requires that the emission
  of the nucleus in the optical strongly dominates over the emission of the
  host galaxy.  Thus, this sample predominantly contains broad line AGN (i.e.\ 
  type-1 AGN), with power law spectra in the optical.
\item[] {\bf AGN:} Galaxies (not point-sources) whose rest-frame color
  properties are consistent with properties of AGN (\pone~$>$~\ponecut, X-ray
  luminosity above $10^{42}$~erg~s$^{-1}$ if X-ray detected).  This
  selection requires that the {\em optical} emission either shows signs of
  both, the emission from the central AGN as well as the emission from the
  underlying host galaxy, or only the latter.  Thus, this sample essentially
  includes Seyfert and LINER types of galaxies, as well as absorption line
  AGN, and we limit it to redshifts of $z\leq1.3$.
\item[] {\bf SF galaxies:} Galaxies whose rest-frame color properties are
  consistent with properties of star forming galaxies (\pone~$\leq$~\ponecut).
  Thus, the emission of these galaxies is dominated by the emission
  originating from regions of substantial star formation. This sample is also
  limited to redshifts $\le1.3$.
\item[] {\bf high-z galaxies}: Galaxies (not point-sources) with redshifts
  beyond $z=1.3$.
\end{itemize}

\subsection{ Star candidates }
\label{sec:stars}

In order to identify star candidates in the VLA-COSMOS matched radio source
sample, we make use of the COSMOS stellar catalog \citep{tasca07}, that was
constructed from the HST/ACS catalog \citep{leauthaud07} using stellar
templates to fit the entire SED of each source.  In \f{fig:ccstars} \ we show
the color-color distribution for $\sim2,000$ objects in the COSMOS field
securely classified as stars (with photometric errors better than 0.05), which
form well defined loci in the broad-band color-color diagrams.
Cross-correlating our matched radio sample with the COSMOS stellar catalog
yields only 2 objects detected in the radio regime that are consistent with
having stellar properties. The color properties of these objects are shown in
\f{fig:ccstars} . Within the error-bars they are consistent with the main
stellar loci. Note, however, the \rmag-\imag\ color excess of one of the star
candidates in the \rmag-\imag\ vs.\ \gmag-\rmag\ color-color diagram (middle
panel in \f{fig:ccstars} ), which suggests consistency with properties of
e.g.\ cataclysmic variables (e.g.\ \citealt{szkody02,szkody03}), or unresolved
binary star systems containing a white dwarf and a late type star (e.g.\ 
\citealt{smo04}).  The best fit stellar templates for these objects were taken
from the PHOENIX library \citep{hauschildt97} and represent dwarfs with
effective temperatures in the range of 4100 to 5000~K and $\log{(g)}$ in the
range of 3 to 3.5. The 1.4~GHz total flux densities for these two objects are 126 and
152~$\mathrm{\mu}$Jy, and the corresponding $i$ band AB magnitudes are 25.34
and 23.28, respectively. It is noteworthy that both objects have IRAC
counterparts, but no associated X-ray emission. We consider these two sources
to be star candidates, however a more detailed analysis (using for example
spectroscopy), which is beyond the scope of this paper, would be needed to
verify this. Such a low fraction of identified stars is consistent with star
detection rates in other deep radio surveys (e.g.\ \citealt{fomalont06}).

The two star candidates in our radio sample form only $\sim0.1\%$ of the
VLA-COSMOS radio sources, and we exclude them from our sample for further
analysis.

\begin{figure}
{\center
\includegraphics[bb = 140 410 400 752, width=\columnwidth]{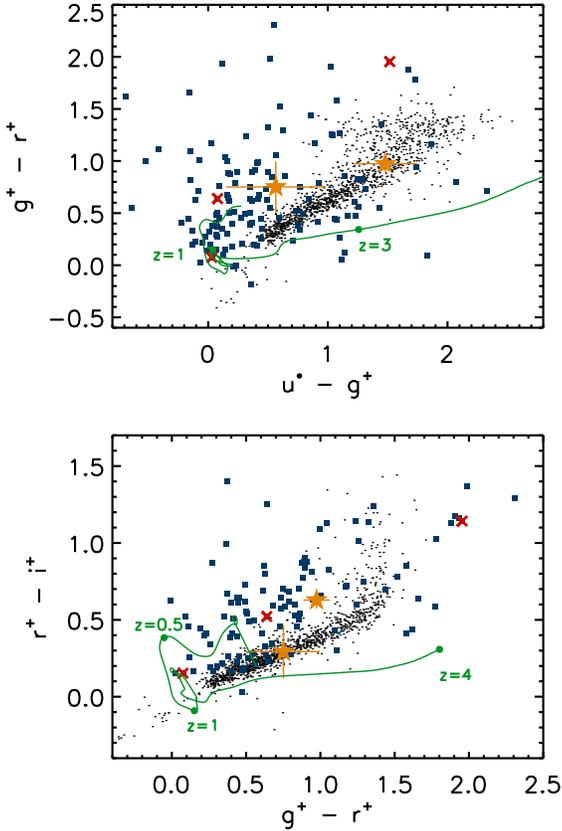} 
\caption{ Color-color diagrams for COSMOS stars and VLA-COSMOS QSOs (see text
  for details). The stars (black dots) form a narrow, well defined, locus in
  each diagram. The star symbols (yellow) show 2 objects detected in the
  VLA-COSMOS radio survey, identified as star candidates (see text for
  details).  VLA-COSMOS objects classified as QSOs by our selection criteria
  (blue squares) are also shown, and they occupy the 'standard' QSO regions in
  these diagrams (see text for details). The crosses (red) show objects
  classified as a QSO by our technique, yet their spectra were identified as
  red galaxies. QSO evolutionary tracks are shown to guide the eye (green
  curved lines).
  \label{fig:ccstars}}}
\end{figure}

\subsection{ Quasi stellar objects }
\label{sec:qso}

\subsubsection{ Identification based on morphology } 

In order to identify QSOs in our matched radio sample we rely on an optical
classification, rather than using X-ray emission, due to the much higher
sensitivity of the observations in the optical ($5\sigma$ sensitivity limit in
the Subaru $i$ band is 26.2; see \citealt{capak07}). For example, if one would
select AGN relying purely on e.g.\ X-ray -- to -- optical flux ratios, which
are generally greater than 0.1 for both broad and narrow line AGN (e.g.\ 
\citealt{maccacaro88, alexander01}), with our optical limit of $i=26$
(corresponding to \rmag\ of $\sim27$) the depth of the X-ray point-source
detection would have to be about 2 orders of magnitude deeper than it
currently is in order to select a complete sample of AGN. Further, a clear
distinction between broad and narrow line AGN would not be possible.  Hence,
here we identify a QSO by requiring that a given source in the matched radio
sample is optically compact. In \f{fig:fwhm} \ we show the fitted $i$ band
FWHM of the sources in the COSMOS photometric catalog \citep{capak07} as a
function of their $i$ band magnitude.  Point-sources (black squares), selected
from the HST/ACS catalog \citep{leauthaud07}, form a locus in this plane,
separated from the area occupied by extended sources. However, the
point-source locus is fairly scattered, especially at faint magnitudes, and
thus makes a single automatic cut at a certain FWHM value inefficient. For
this reason, we classify sources within the FWHM range of $1.85''-2.05''$ as
QSOs only if their optical HST morphology was visually confirmed to be
'point-source dominated'. However, we consider {\em all} sources below 
FWHM of $1.85''$ to be QSOs.  We further supplement this sample with 12 objects
that were classified as point-sources in the HST/ACS catalog, but do not
satisfy the above criteria.

% 7
\begin{figure}
{\center
\includegraphics[bb = 14 14 339 281, width=\columnwidth]{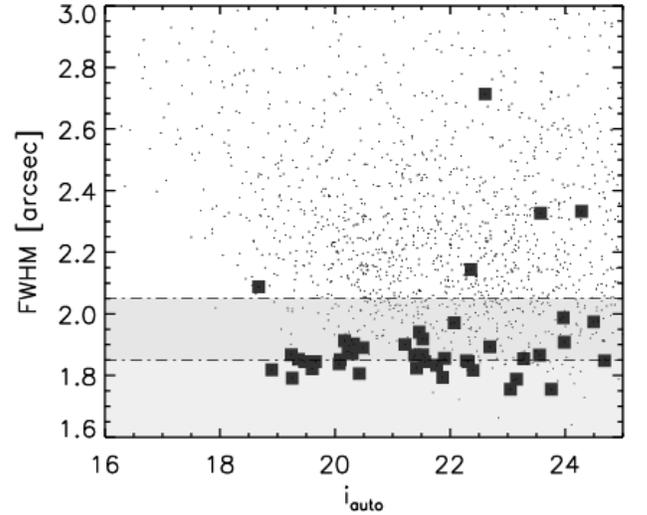}
\caption{ FWHM of the optical counterparts of the matched radio sources (small
  dots) as a function of $i$ band magnitude, taken from the COSMOS photometric
  catalog \citep{capak07}. Grey squares represent the HST/ACS point sources in
  the sample. Unresolved sources form a separated locus in this diagram.  We
  identify QSOs by requiring that a source is optically compact by choosing an
  absolute separation value of $\mathrm{FWHM}\lesssim 1.85''$ (below lower
  dash-dotted line; light-grey shaded area). In the range of
  FWHM~$\in(1.85,2.05]''$ (between upper and lower dash-dotted lines;
  dark-grey shaded area) we identify objects as QSOs only if they were
  visually confirmed to be point-source dominated (see text for details).
  \label{fig:fwhm}}}
\end{figure}

In summary, out of \no\ objects we identify \noQSO\ (i.e.\ $11.5\%$) as QSOs.
In \f{fig:ccstars} \ we show their broad-band (\umag,\gmag,\rmag,\imag)
color-color properties. As expected, the non-stellar emission of the selected
objects confines them to regions typical for QSOs, which are separated from
the main stellar loci in these diagrams (e.g.\ \citealt{brusa07, richards02}).
A minor fraction of these objects lie on the stellar loci. However, in the
$BzK$ diagnostic diagram, which is an efficient tracer for stars
\citep{daddi04} these sources are offset from the stellar locus, verifying
their non-stellar nature.
%For comparison the
%expected color-color track for QSOs from a redshift of $0$ to $4$ is also
%shown in \f{fig:ccstars} .

%\input{QSO_Xray.tex}

As AGN dominated systems usually have soft X-ray to optical flux ratios in the
range of about $0.1$ to $10$ (e.g.\ \citealt{maccacaro88, alexander01}), we
can use the X-ray to optical flux properties of the identified QSOs to further
test our selection criteria. 43 objects in our QSO sample were detected as
X-ray point-sources, and their X-ray to optical flux ratios are consistent
with the expected values. The median \rmag\ magnitude for these sources is
21.3. For the remaining QSO candidates, that were not detected in the X-rays,
the median \rmag\ is 24.4.  Therefore, these sources are also consistent with
the expected X-ray to optical flux ratios, however beyond our X-ray
point-source detection limit ($10^{-15}$~erg~cm$^{-2}$~s$^{-1}$ in the soft
band). We conclude that the independent analysis of the X-ray properties of
our selected QSOs verifies the validity of our selection.

\subsubsection{Spectroscopic verification}
A sub-sample of 31 objects of the \noQSO\ previously identified QSOs have
available optical spectroscopy with secure classifications \citep{trump07,
  prescott06, colless01,schneider05}, and only 3 of these objects were
classified as red galaxies \citep{trump07}, while all the others have AGN
classifications. The 3 galaxies classified as ellipticals were identified as
QSO candidates by our method based on visual/morphologic classification, which
suggests the presence of dominating nuclear emission. The color properties of
two of theses objects (see red crosses in the top panel in \f{fig:ccstars} )
are also consistent with the color properties of quasars.\footnote{For
  example, in the \umag-\gmag\ vs.~\gmag-\rmag\ color-color diagram (top panel
  in \f{fig:ccstars} ) red galaxies would occupy the upper right quadrant (see
  e.g.\ \citealt{strateva01}). } Thus, we conclude that the selected QSO
sample is not significantly ($\lesssim10\%$) affected by contamination of
non-QSO objects.

In order to assess the completeness of the selected QSO sample, we search for
objects that are spectroscopically classified as QSOs and 'missed' by our
\CLASSmethod. Our criteria yielded \noQSO\ objects classified as QSOs in the
matched radio sample, and in the remainder of the sample (i.e.\ the \noNoQSO\ 
sources that were not classified as star candidates or QSOs) 
spectroscopic classifications are available for 397 objects. Out of these, 9
were spectroscopically classified as QSOs. Two SDSS examples, for which COSMOS
HST/ACS imaging is available, are shown in \f{fig:noq1s} .  They obviously
show extended optical emission, and a substantial light component arises from
the host galaxy itself. The median redshift of these 9 objects is only 0.4.
It is noteworthy that all of these objects have X-ray point source detections,
and all except one have X-ray luminosities higher than
$10^{42}$~erg~s$^{-1}$. Therefore these galaxies will be selected into our AGN
class, hence not contaminating the SF galaxy sample (see \s{sec:RF} ).
As the spectroscopic sub-sample fairly represents the full matched radio
sample (see \f{fig:pophisto} ), we conclude, based on the above analysis, that
the sample of identified QSOs is about 80\% complete. As expected, the
incompleteness is mostly due to relatively low redshift, low-luminosity AGN.

\begin{figure}
{\center
  \includegraphics[bb = 14 14 329 240, width=\columnwidth]{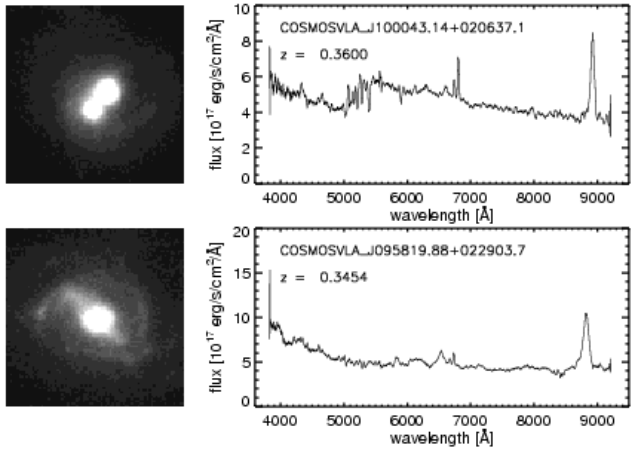}
\caption{ HST/ACS stamps ($4''$ on the side; left panels) and the
  corresponding SDSS spectra (smoothed using $20$~\AA\ wide bins; right
  panels) for two sources that were spectroscopically classified as QSOs, but
  missed by our QSO selection method. The morphology of these sources is
  clearly extended (in particular the source in the top panel is possibly a
  merging system with double nuclei visible, and extending beyond the
  displayed $4''$).
\label{fig:noq1s}}}
\end{figure}

\subsection{ Star forming and AGN galaxies  } 
\label{sec:RF}

%\subsubsection{ Identification based on the \RFmethod }

In the previous sections we have identified 2 star candidates and \noQSO\ QSOs
in the matched radio sample. We will refer to the \noNoQSO\ remaining sources
in the matched radio source sample as the 'galaxy sample'. Before applying the
\RFmethod\ to our VLA-COSMOS galaxies in order to separate SF from AGN
galaxies, we restrict the galaxy sample to \noGAL\ galaxies with redshifts
$\leq1.3$, as a) the photometric redshifts are less reliable beyond this
redshift, and b) the library of BC03 model spectra that we use for the SED
fitting may not be appropriate for fits beyond this redshift as the
distribution of priors was set to optimally match this intermediate redshift
range. Hereafter, we will call the sample of \noUNK\ galaxies with redshifts
greater than 1.3 high redshift (high-z) galaxies.
 
We perform an SED fit using GOSSIP (as described in \s{sec:synthesis} ) for
each of the \noGAL\ objects in the matched radio 'galaxy' sample out to
$z=1.3$. The distribution of the rest-frame color \pone\ for these galaxies is
shown in \f{fig:p1} . The distribution is very similar to that of the local
sample (see top panel in \f{fig:p1sdssnvss} ) with a peak at \pone$\sim0.4$
(AGN) and a prominent tail towards bluer values (SF galaxies). We inspected
the behavior of the median value of the synthesized \pone\ color for the
entire $z\leq1.3$ galaxy sample as a function of redshift, and we found no
significant evolution in the median color.  We reached the same conclusion
analyzing the median \pone\ colors of the SF and AGN sub-populations. This
implies that a fixed cut in the color can safely be applied to the entire
galaxy population out to $z=1.3$.

\begin{figure}
{\center
  \includegraphics[bb = 14 14 340 231, width=\columnwidth]{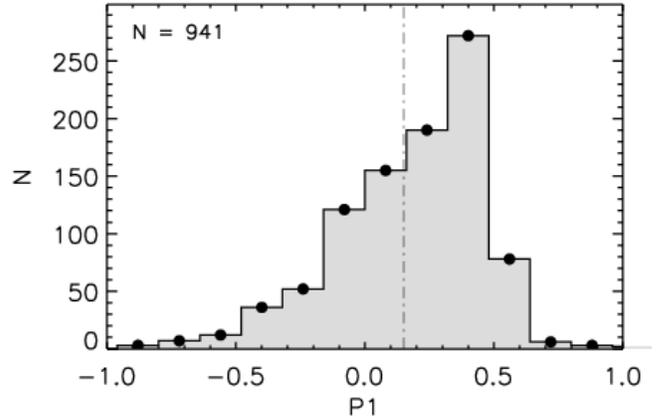}
 \caption{ The distribution of the synthesized rest-frame \pone\ color,
   corrected for the systematics (see bottom panel in \f{fig:p1sdssgossip} ),
   for the objects in the matched radio {\em galaxy} sample with $z\leq1.3$.
   Note, that the expected distribution from our local sample, showing a
   prominent tail towards bluer values of \pone, is reproduced. The
   dash-dotted line indicates our chosen boundary for separating star forming
   from AGN galaxies.
   \label{fig:p1}}}
\end{figure}

The SED fitting was performed via a $\chi^2$ minimization procedure. The
median value of the reduced $\chi^2$ of the SED fits, computed using the best
fit model spectrum, is $0.6$ with an interquartile range of $2$. Only 10\% of
the fitted objects have reduced $\chi^2$ values above 5, and only 5\% above
10. A visual inspection suggests that these galaxies are either nearby
galaxies, which are resolved and often saturated in the Subaru $i$ band, or
QSO contaminants. While the latter predominantly have blue \pone\ colors, the
synthesized \pone\ color for the first class of galaxies still appears to be a
valid tracer for the SF/AGN separation, and therefore we do not reject them
from the sample.
In order to select SF and AGN galaxies we require that the synthesized \pone\ 
color is $\leq$~\ponecut\ and $>$~\ponecut, respectively. However, to improve
our selection at this point we make use of the X-ray properties in the soft
band of the 114 galaxies that were detected as X-ray point sources. Namely, if
the soft X-ray luminosity of an object is greater than $10^{42}$~erg~s$^{-1}$
we consider it to be an AGN, regardless of its \pone\ color.  Note that this
criterion is expected to reduce the contamination of the SF sample by objects
with blue rest-frame colors, such as QSOs missed by our selection. Out of
  the 114 X-ray detected sources, 77 have X-ray luminosities greater than the
  above given value, and 37 out of these 77 have a synthetic \pone\ color
  $\leq$~\ponecut . In summary, our selection yields \noSF\ SF and \noAGN\ AGN
galaxies in our matched radio galaxy sample with $z\leq1.3$.  We analyze these
galaxies further in \s{sec:resultsRF} .

\section { Comparison with other selection methods }
\label{sec:comparison}
In this Section we compare our \CLASSmethod, that we have applied to an
intermediate redshift population, with other classification methods used for
both local and intermediate redshift populations in the literature
\citep{lacy04, stern05, best05}. We also study the 24\mic\ properties of our
radio sources, and their correlation to the 1.4~GHz emission.

\subsection{3.6-8~\mic\ color -- color diagnostics}
QSOs, whose UV to NIR continuum is dominated by a power law, tend to be redder
than other types of galaxies in the MIR. Hence, they occupy a distinct region
in MIR color space, and several color-color criteria were suggested for their
selection \citep{lacy04, stern05}. In \f{fig:lacystern} \ we compare our
\CLASSmethod\ with those proposed in the MIR using a sub-sample of the matched
radio sources that were also detected with IRAC ($\sim90\%$ have IRAC
counterparts; see \s{sec:CCradio-opt-spitzer} ).  We indicate the QSO (dots),
AGN (thin contours) and star forming (thick contours) galaxies selected using
our method. The dashed lines in the top and bottom panel in \f{fig:lacystern}
\ show the color-color criteria proposed by \citet{lacy04} and
\citet{stern05}, respectively, for the selection of broad-line AGN. As
expected, the majority of objects selected as QSOs by our method falls within
this region, reassuring the efficiency of the \CLASSmethod\ presented here.
There are several QSO candidates outside these regions, which is not
surprising as the suggested 'quasar regions' do not select a 100\% complete
sample of QSOs, and a certain amount of outliers is expected (see
\citealt{stern05} for a discussion of this point). In \s{sec:qso} \ we have
inferred that our selected sample of QSOs is not significantly contaminated by
different types of objects, which is affirmed by this independent analysis.

\begin{figure}
{\center
  \includegraphics[bb = 14 14 334 279, width=\columnwidth]{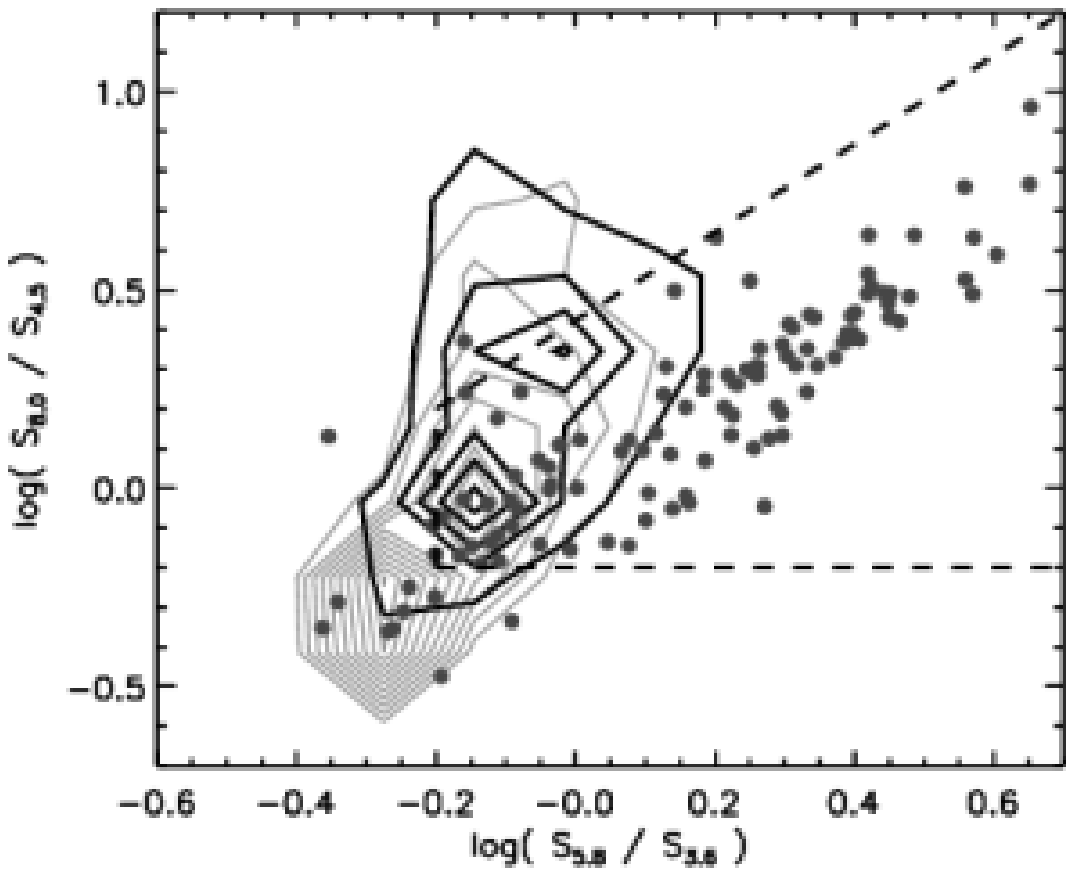}
\includegraphics[bb =  14 14 335 279, width=\columnwidth]{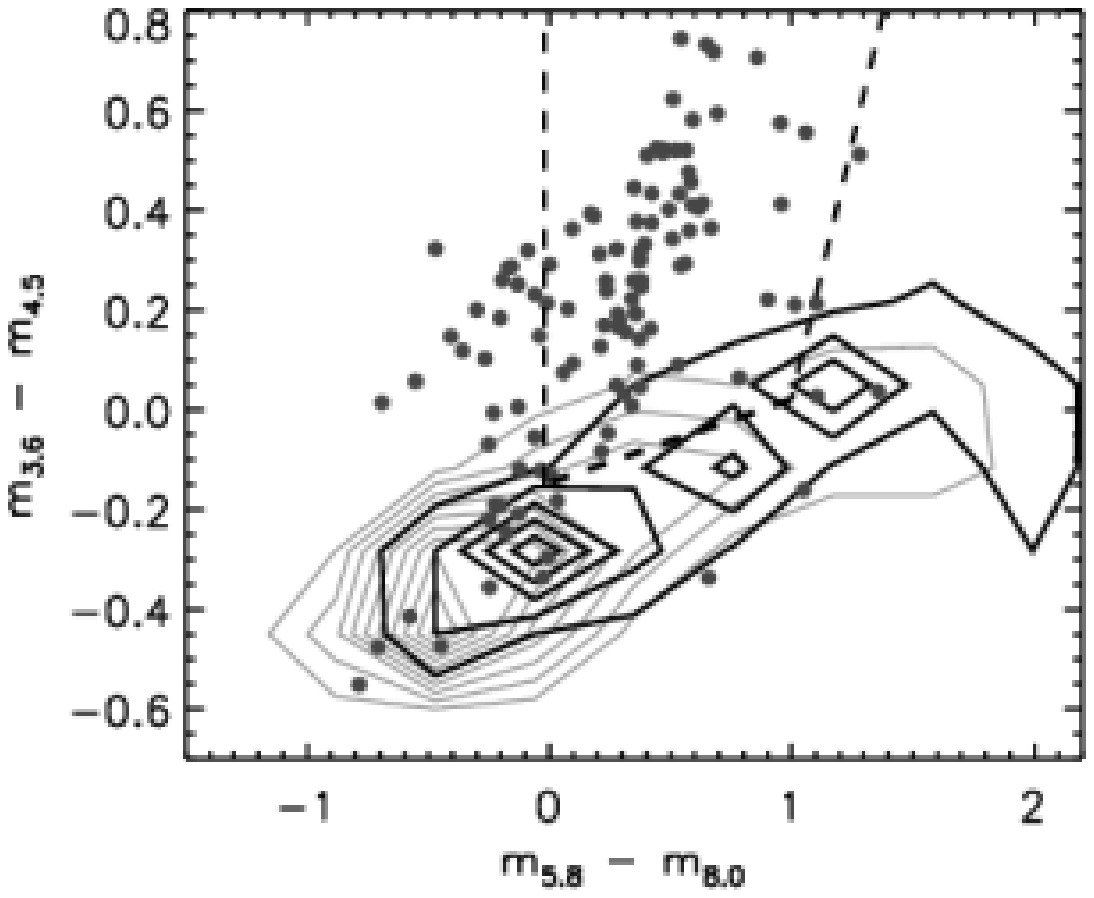}
 \caption{ IRAC color-color diagrams in two representations (top and bottom
   panel) used as diagnostic tools for the separation of QSOs (broad line AGN)
   from 'normal' galaxies \citep{lacy04, stern05}. %The complete IRAC-COSMOS
   %sample is shown in a background grey-scale density representation.  
   Filled circles display the 122 QSOs selected by our morphologic method that
   have  IRAC detections; thin and thick contours show the 579 AGN and 322
   star forming galaxies separated by our \RFmethod, respectively, that have
   IRAC counterparts. The contour levels in both panels are in steps of 7
   starting at 7. The dashed lines in the top and bottom panels show the
   empirically selected regions for identifying broad-line AGN proposed by
   \citet[top panel]{lacy04} and \citet[bottom panel]{stern05}, respectively.
   \label{fig:lacystern}}}
\end{figure}

\citet{stern05} showed that at redshifts of $\lesssim1$ galaxies span a large
range in the $m_{5.8} - m_{8.0}$ color, which is consistent with the
horizontal extent of our selected star forming and AGN galaxies (see bottom
panel in \f{fig:lacystern} ).  However, typical low luminosity AGN and
starburst galaxies cannot be clearly divided using these diagnostic diagrams
(see also color evolutionary tracks in \f{fig:stern4matched} ). Nonetheless,
elliptical galaxies (which correspond to our class of absorption line AGN)
tend to occupy the bottom left regions in both diagrams, and close to these
regions the distributions of our identified AGNs peak. On the other hand, the
peak of the distribution of our selected star forming galaxies in these
diagrams is clearly displaced from the one for AGN. We would like to
  stress that this independently confirms that indeed two different
  populations of galaxies have been selected. The differences in the MIR
  properties of our SF and AGN galaxies affirm the strength of the \RFmethod.

%The differences in the MIR properties of our SF and AGN
%galaxies are consistent with the absence of star formation in otherwise normal
%galaxies, which have been classified as AGN on the basis of our \RFmethod.

Further, \citet{stern05} showed that narrow line AGN appear spread out in both
the QSO and galaxy regions, which is also a result of our selection method
[note that the selected AGN are present in both regions]. The last point we
want to stress is that the star forming galaxy locus in these diagrams is also
consistent with the expected colors, as a 'contamination' by star forming
galaxies of the QSO locus is expected, especially close to the boundary. In
summary, the \CLASSmethod\ presented here agrees remarkably well with the
expected properties of QSOs, AGN and star forming galaxies at intermediate
redshifts in the MIR range encompassing 3.6-8~\mic.

\subsection{ 24~\mic\ properties: The 24~\mic\ -- radio correlation  } 

A tight mid-IR (as well as far- and total- IR) -- radio correlation is
expected for star forming galaxies, while 'radio-loud' AGN are expected to
strongly deviate from it (e.g.\ \citealt{condon92,bell03,appleton04}). The
60~\mic\ -- radio correlation for low-luminosity AGN was studied by
\citet{obric06} in the local universe. Based on a selection utilizing the BPT
diagram they have shown that also low -luminosity AGN follow a tight FIR --
radio correlation, however with a slightly different slope and a larger
scatter than SF galaxies.
%They concluded that SF galaxies in the local
%universe appear to show $\sim20\%$ more radio emission relative to the FIR
%than AGN. 
In this section we investigate the 24~\mic\ -- radio correlation for
our selected SF and AGN galaxies. In particular, if our SF/AGN separation
method is successful, then a difference in the 24~\mic\ compared to 20~cm
properties is expected to be seen for the two populations.

Our \RFmethod\ has identified \noSF\ SF galaxies. Out of these $82\%$ (280)
were detected at 24~\mic\ with a signal to noise $\geq 3$. On the other hand,
out of \noAGN\ selected AGN only $44\%$ (267) have a MIPS 24~\mic\ detection
with $\mathrm{S/N} \geq 3$.  In \f{fig:lq} \ we show the 24~\mic\ vs. 1.4~GHz
luminosity (top panel) for our SF and AGN galaxies, where the 24~\mic\ 
data was not k-corrected. A correlation between the two luminosities exists
for both types of objects detected at 24~\mic, although on average for a given
$\mathrm{L_\mathrm{1.4GHz}}$ the 24~\mic\ luminosity is slightly lower for AGN
than for SF galaxies (see also below). For the SF and AGN galaxies that were
not detected at 24~\mic\ we have computed upper limits of the 24~\mic\ 
luminosity using the detection limit of the S-COSMOS MIPS shallow survey which
is 0.3~mJy. These limiting luminosities are also shown in \f{fig:lq} . Note
that for AGN galaxies, as $56\%$ of them are not detected at 24~\mic, the
scatter in the correlation is significantly increased by these objects.

\begin{figure}
{\center
  \includegraphics[bb =   14 14 343 416, width=\columnwidth]{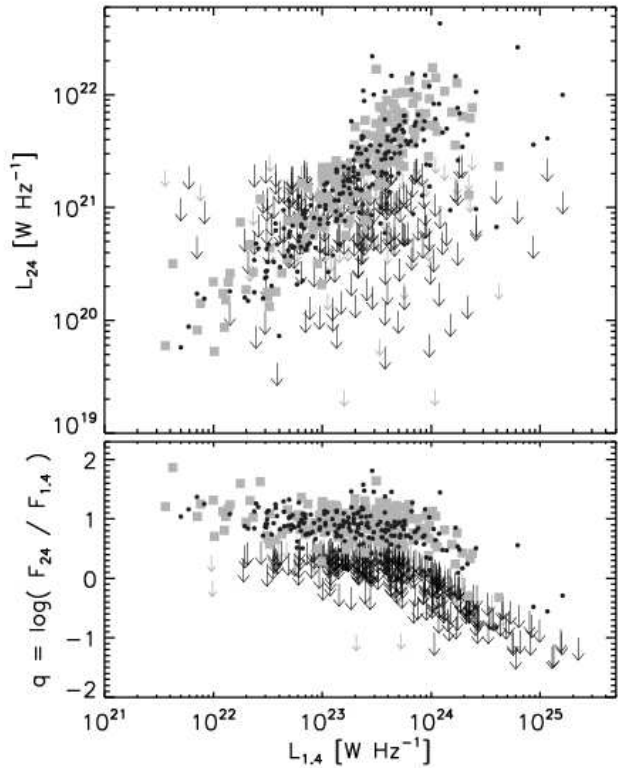}
 \caption{ {\em Top panel}: 24~\mic\ luminosity vs. 1.4~GHz luminosity for our
   selected SF (grey squares) and AGN (black dots) galaxies. Upper 24~\mic\ 
   luminosity limits for the radio SF ($18\%$) and AGN ($56\%$)
   galaxies that were not detected at 24\mic\ are indicated by grey and black
   arrows, respectively.  {\em Bottom panel}: The $q$ parameter, defined as the
   logarithm of the 24~\mic\ to 20~cm flux ratio, as a function of 20~cm
   luminosity for SF (grey squares) and AGN (black dots) galaxies. Upper
   limits are indicated by arrows, as in the top panel. Note the much larger
   scatter in $q$ for AGN than for SF galaxies.
   \label{fig:lq}}}
\end{figure}

To quantify the correlation, we derive the classical $q$ parameter (e.g.\ 
\citealt{condon92}) as the logarithm of the 24~\mic\ to 1.4~GHz observed flux
ratios.  This parameter essentially measures the slope of the correlation, and
in the bottom panel in \f{fig:lq} \ we show it as a function of
$\mathrm{L_\mathrm{1.4GHz}}$ for our SF and AGN galaxies, with indicated upper
limits (derived as described above).  The $q$ parameter seems to show a
decreasing trend with increasing radio luminosity. However, this trend is
dominated by the objects that have only estimated upper limits, and therefore
may be mimicked by the flux limits of the samples. A more detailed analysis of
this issue is beyond the scope of this paper.

\begin{figure}
{\center
  \includegraphics[bb = 54 360 486 722, width=\columnwidth]{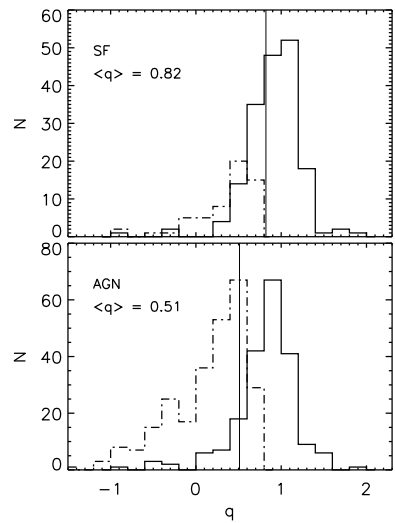}
 \caption{ The distribution of the $q$ parameter (see also \f{fig:lq} ) for SF
   (top panel) and AGN (bottom panel) galaxies. In both panels the solid
   histograms show the distribution of the 1.4~GHz sources detected at
   24~\mic\, while the dash-dotted histograms represent the distribution of
   the upper limits of $q$ obtained for the radio sources that were not
   detected at 24~\mic\ (see text for details).  The solid vertical line in
   each panel designates the median value of the entire distribution, also
   listed in the top left in each panel.
   \label{fig:qhisto}}}
\end{figure}

The distribution of the $q$ parameter for SF and AGN galaxies is shown in
\f{fig:qhisto} , for galaxies detected at 24~\mic\ and those which only
have upper limits. The median $q$ value for SF galaxies is $0.82\pm0.05$ with
a scatter of $\sim0.3$ when all objects (also the upper limits) are taken into
account. On the other hand, the median $q$ value for the AGN galaxies is
$0.51\pm0.02$, significantly lower than for SF galaxies. We also find a larger
spread in q ($\sim0.4$) for the AGN population. Note, however, that the spread
quoted here should be considered somehow tentative, especially for AGN, as the
{\em exact} $q$ values for the fraction of objects not detected at 24~\mic\ 
are not known.  Nonetheless, this does not affect the estimates of the median
values.  Our $q$ parameter derived for SF galaxies is remarkably consistent
with the one inferred by \citet{appleton04} at 24~\mic. Combining Spitzer --
MIPS and VLA observations in the First Look Survey with optical spectroscopy,
they have found a $q$ value of 0.84 with a spread of 0.28 (with no
k-corrections applied). They have also shown that the FIR luminosities of AGN
tend to be lower for a given radio luminosity, consistent with our findings
here.

Finally, in \f{fig:qz} \ we show the $q$ parameter as a function of redshift
for our SF and AGN galaxies. $q$ does not depend on redshift, both for SF and
AGN galaxies, implying that the MIR -- radio correlation with the same slope is
valid out to high redshifts ($z\sim1.3$). This result is again consistent with
those presented in \citet{appleton04}. 

In summary, the above results have shown that the 24~\mic\ -- radio correlation
has different properties for our selected SF and AGN galaxies, which verifies
the efficiency of our \RFmethod.

\begin{figure}
{\center
  \includegraphics[bb = 14 14 284 208, width=\columnwidth]{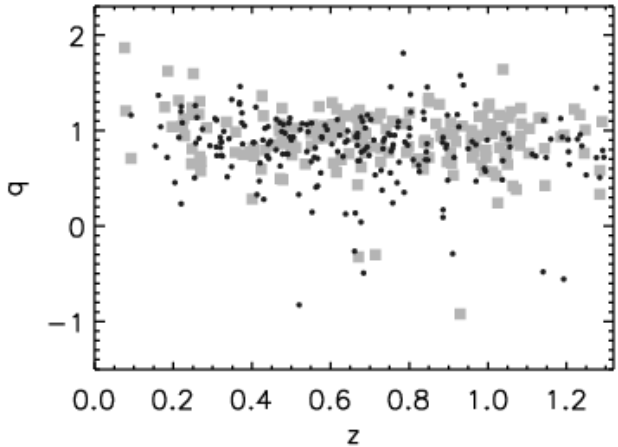}
 \caption{ $q$ as a function of redshift for SF (grey squares) and AGN (black
  dots) galaxies. Note that $q$ does not depend on redshift, implying that the
  24~\mic -- radio correlation holds out to high redshift (see text for
  details). 
   \label{fig:qz}}}
\end{figure}

\subsection{ Selection based on optical spectroscopic properties, radio
  luminosity and stellar mass }

\citet{best05} defined a sample of $\sim3000$ local ($0.01<z<0.3$) galaxies
from the SDSS DR2 ``main'' spectroscopic sample matched with sources above
5~mJy detected in the NVSS survey. They further divided the radio -- optical
source sample into AGN and SF galaxies making use of the galaxies' location in
the plane spanned by the $4000$~\AA\ break [\dn] and radio luminosity
[$\mathrm{L}_{1.4\mathrm{GHz}}$] normalized by stellar mass [$\mathrm{M}_*$].
In \f{fig:best} \ we compare our \RFmethod\ with the one utilized by
\citet{best05} in the local universe.  The \dn~vs.~$\log{\left(
    \mathrm{L}_{1.4\mathrm{GHz}} / \mathrm{M}_* \right)}$ distribution for all
galaxies in the matched radio source sample is shown in \f{fig:best} .  The
1.4~GHz luminosity for these galaxies was derived, and \dn\ and $\mathrm{M}_*$
from the best fit template from the SED fitting (see \s{sec:p1derivation} ).
The average errors are indicated.  The dashed line corresponds to the
separation between SF and AGN proposed by \citet{best05}, and the two types of
symbols designate the SF (squares) and AGN (dots) galaxies identified by our
\RFmethod.  We want to note that Best et al.\ calibrated their separation
method using a slightly different selection of objects in the BPT diagram (see
their Fig.~9) with respect to the one we use here.  Therefore, a perfect
correspondence between our and the Best et al.\ method is not to be expected,
even if our derived quantities were absolutely accurate.  The area in the
\dn~vs.~$\log{\left( \mathrm{L}_{1.4\mathrm{GHz}} / \mathrm{M}_* \right)}$
plane where the major disagreement is expected, due to the different
selections in the BPT diagram, is in the range of $1.4<$\dn~$<1.6$, and
$11<\log{\left(\mathrm{L}_{1.4\mathrm{GHz}} / \mathrm{M}_* \right)}<12$. This
is the region where a larger fraction of Seyfert and LINER galaxies is located
(see Fig.~9 in \citealt{best05}), and, different from Best et al., we define
these galaxies exclusively as AGN. In this region in \f{fig:best} \ we indeed
see the largest disagreement between the two classifications. Further, the
existence of objects with \dn~$<1.3$ that we classify as AGN is not
surprising, but it rather reflects the dual properties of composite objects,
which in this case were classified as AGN by our \RFmethod.  We also want to
note that the average error in the synthesized \dn\ [derived from comparison
with the spectroscopic and synthetic \dn\ in the local sample] is fairly
large, and thus prevents a more detailed comparison between the two selection
methods.  Overall, given the error bars and the difference in the basic
selections of the two methods, as well as the fact that our \dn\ values are
{\em not} spectral measurements on the data, but values taken from the best
fit template, we conclude that our \RFmethod\ agrees well with the one
proposed by \citet{best05} in the local universe.

In summary, our \RFmethod\ for separating SF from AGN galaxies agrees well
with other selection schemes, proposed in the literature, which are based
both on MIR colors and optical spectroscopic diagnostics.

\begin{figure}
{\center
\includegraphics[bb = 14 14 334 280, width=\columnwidth]{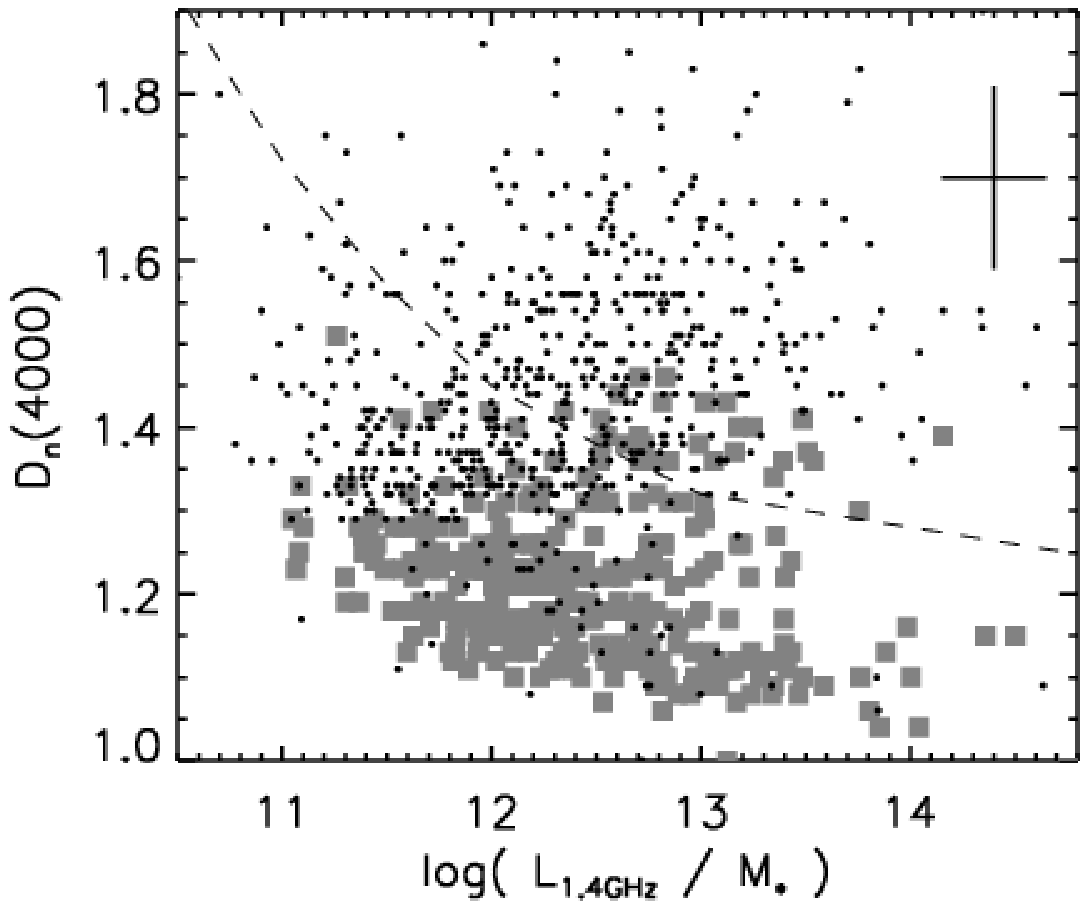}
 \caption{ The $4000$~\AA\ break, \dn, vs.\ 1.4~GHz radio luminosity
   normalized by stellar mass, $\mathrm{L}_\mathrm{1.4GHz}/\mathrm{M}_*$ for
   galaxies in the matched radio sample.  $\mathrm{M}^*$ and \dn\ were
   synthesized via SED fitting by GOSSIP.  The black and grey symbols show
   objects classified by our \RFmethod\ as AGN and star forming galaxies,
   respectively. The dashed line corresponds to the separation proposed by
   \citet{best05} for local galaxies. The average errors of the synthesized
   quantities are shown in the top right.  A good consistency between the two
   selection methods exists, given the large error bars, as well as a slightly
   different selection rational used here and in Best et al.\ (2005a; see text
   for details).
   \label{fig:best}}}
\end{figure}

\section { Discussion: The composition of the faint radio population }  
\label{sec:resultsRF}

In previous Sections we have presented, tested, and discussed in detail the
photometric \CLASSmethod\ which we used to separate the matched radio source
sample into stars, QSOs, star forming, AGN and high-z galaxies. In this
Section we discuss the properties of the 'population mix' in the VLA-COSMOS
survey: In \s{sec:Lzdistrib} \ we describe the redshift and luminosity
distributions of the selected SF and AGN galaxies, and in \s{sec:popmix} \ we
study the contribution of different source types to the sub-mJy radio
population. We show, based on the matched radio sample, as well as on the
remaining radio sources with no optical counterparts (brighter than $i=26$),
that star forming galaxies do not dominate the sub-mJy sources, but that the
majority of these sources is rather comprised of AGN and QSOs.

\subsection{ The redshifts and luminosity distributions of SF and AGN galaxies
  out to $z=1.3$ } 
\label{sec:Lzdistrib}

The \RFmethod\ yielded \noSF\ star forming and \noAGN\ AGN galaxies out to
redshifts of $1.3$.
% (\noUNK\ galaxies beyond $z=1.3$ were classified as UNKs).
In the top panel in \f{fig:zsfagn} \ we show the redshift distribution for
these galaxies using redshift bins of 0.217 in width. We use such wide
redshift bins to assess the {\em average} properties of the radio population,
reducing the effects of fluctuations due to the strong and narrow
overdensities which are known to exist in the COSMOS field
\citep{scoville07,finoguenov07,smo07b}. Poisson errors are indicated for each
bin.  The deficit of galaxies at the low-redshift end reflects the relatively
small comoving volume sampled by the 2\sqdeg\ area of the COSMOS field at
these redshifts. The decline in the number of sources at the high-redshift
end, on the other hand, reflects the detection limit of the VLA-COSMOS survey.
The redshift distribution of the number of star forming galaxies seems to be
more uniform than the one for AGN, in particular the relative number of star
forming galaxies compared to AGN rises at higher redshifts ($z\sim1$). This
may be explained by the relatively high number density of ULIRGs expected at
these redshifts \citep{le floc'h05,caputi07} in conjunction with the
VLA-COSMOS detection limit which at these redshifts allows to sample only
radio luminosities larger than $3\times10^{23}$~WHz$^{-1}$ (see \f{fig:limits}
\ below). Further, as the comoving volume surveyed at $z\sim1$ is larger than
the one surveyed locally, the probability of detecting a ULIRG is also
increasing at these redshifts.  Effects of cosmic variance as a function of
redshift cannot be excluded, however they should be smaller than for other
deep radio surveys that typically probe significantly smaller areas.  An
  increase of the AGN fraction at $z\sim0.7$ is discernible. It possibly
  arises due to the dense large scale structure component in the COSMOS field
  at this redshift \citep{scolss, guzzo07}.

\begin{figure}
{\center
\includegraphics[bb =  18 20 330 198, width=\columnwidth]  {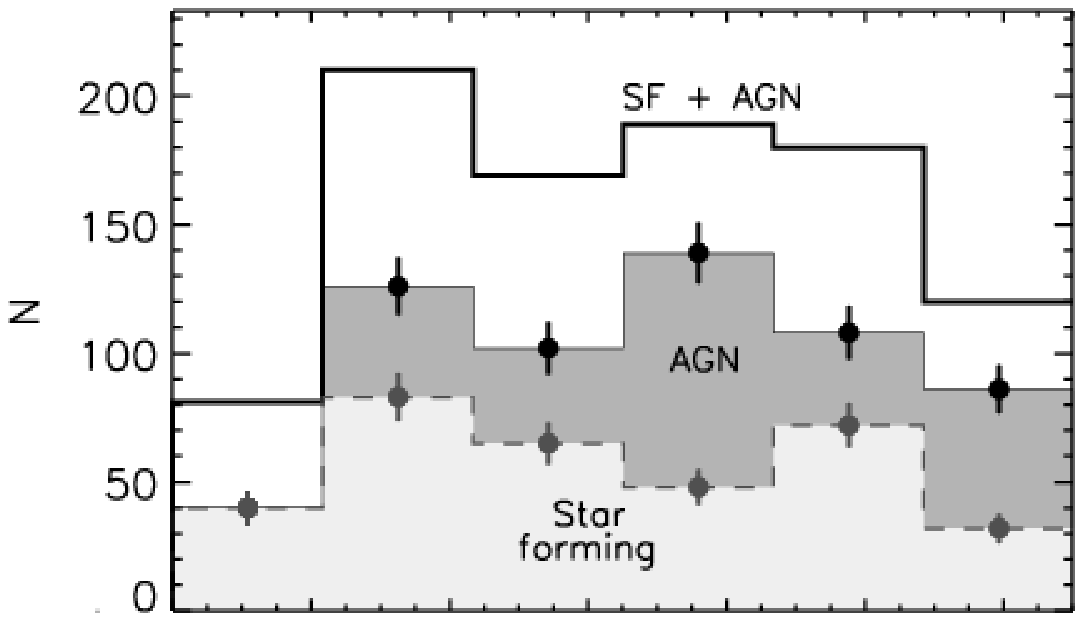}
\includegraphics[bb =  14 14 329 232, width=\columnwidth]  {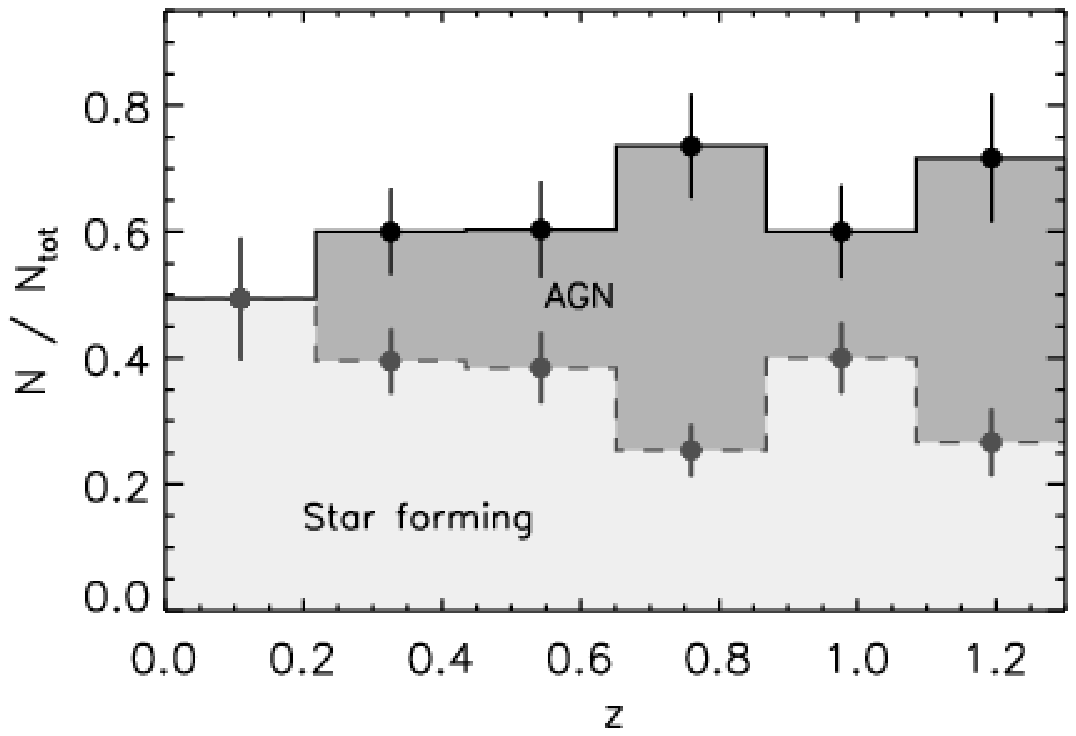}
 \caption{ {\em Top:} Redshift distribution of i) the galaxies in the matched
   radio sample out to $z=1.3$ (thick line), ii) the \noSF\ 
   selected star forming galaxies (light-grey shaded histogram) and iii)
   \noAGN\ identified AGN (dark-grey shaded histogram). {\em Bottom:} The
   fractional distribution of the SF and AGN galaxies compared to the total
   number of galaxies in the matched radio sample out to $z=1.3$ as a function
   of redshift. In both panels Poisson errors are indicated.
   \label{fig:zsfagn}}}
\end{figure}

In the bottom panel in \f{fig:zsfagn} \ we show the fractional contribution of
the SF and AGN galaxies to the $z\le1.3$ matched radio population, as a
function of redshift.  On average, we find that the mean fractional
contribution of SF and AGN galaxies to the $z\leq1.3$ matched radio population
is $(34\pm14)\%$ and $(66\pm14)\%$, respectively.  This is strikingly similar
to the relative numbers of SF and AGN galaxies in the local Universe. Namely,
if we apply the adopted \pone\ color cut to the SDSS/NVSS galaxy sample (see
\s{sec:locmet} ), we find that $\sim32\%$ of the galaxies are star forming,
and $\sim68\%$ are AGN. If, as shown by the tests described in \s{sec:locmet}
, our \RFmethod\ can reliably be applied also to high redshift galaxies, then
the similarity of the SF and AGN fractions suggests that the two populations
have similar evolutionary properties out to $z\sim1.3$.

In \f{fig:limits} , we show the 1.4~GHz luminosity as a function of redshift
for the selected SF and AGN galaxies out to $z=1.3$. We also indicate the
expected luminosity ranges for Milky Way type galaxies, LIRGs, ULIRGs and
HyLIRGs, which were derived using the total IR -- radio correlation
\citep{bell03}. It is noteworthy that the majority of galaxies with
luminosities typical for HyLIRGs ($\mathrm{L_{IR}}>10^{13}$~\lsun) was
classified by the \RFmethod\ as AGN, consistent with the expected properties
of these galaxies (e.g.\ \citealt{veilleux99, tran01}).  This point is seen
more clearly in \f{fig:lsfagn} , where we show the distribution of the 1.4~GHz
luminosity for the selected star forming and AGN galaxies.  The median
luminosities are $\sim1.6\times10^{23}$~W\,Hz$^{-1}$ and
$\sim3.2\times10^{23}$~W\,Hz$^{-1}$ for SF and AGN galaxies, respectively.
Although the median luminosities of the two populations are different only by
a factor of 2 (note that this is drawn from a luminosity distribution of
  a flux limited sample), there are some significant differences at both high
and low radio luminosity.  At high luminosities there is the strong decline of
the number of SF galaxies with luminosities above $\sim10^{24}$~W\,Hz$^{-1}$,
while AGN show an extended tail towards the brightest 1.4~GHz luminosities.
Such a behavior is consistent with results from local studies, which suggested
that 'normal'\footnote{'Normal' galaxies, in terms of radio properties, are
  broadly defined as galaxies whose radio emission is not powered by a
  super-massive black hole. These galaxies are a mix of spiral, dwarf
  irregular galaxies, peculiar and interacting systems, as well as E/S0
  galaxies with ongoing star formation (see \citealt{condon92} for a review).}
galaxies tend to have
$\mathrm{L}_{1.4\mathrm{GHz}}\lesssim10^{24}$~W\,Hz$^{-1}$ (e.g.\ 
\citealt{condon92}).
%, which is one order of magnitude below the FRI -- FRII
%\citep{fr74} break. 
It is noteworthy, that our SF and AGN galaxies were identified completely
independently from their radio luminosity, yet their 1.4~GHz luminosities
match the expectations based on local studies.  At low luminosity ($\lesssim
2\times 10^{22}$~W\,Hz$^{-1}$; below the typical LIRG radio luminosity) the
fraction of SF galaxies increases and the numbers of SF and AGN galaxies are
similar to each other.

\begin{figure}
{\center
\includegraphics[bb =  14 14 335 234, width=\columnwidth] {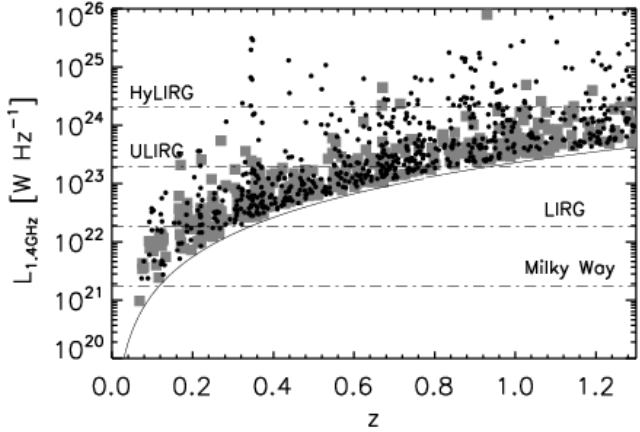}
 \caption{ 1.4~GHz luminosity as a function of redshift for the \noSF\ selected
   star forming galaxies (grey squares) and \noAGN\ AGN (black dots). The
   horizontal (dot-dashed) lines correspond to 1.4~GHz luminosities typical
   for various classes of galaxies, obtained using the total IR -- radio
   correlation \citep{bell03}. The solid curved line corresponds to the
   VLA-COSMOS $5\sigma$ limit of $\sim50~\mathrm{\mu Jy}$. Note also that the
   VLA-COSMOS survey is sampling the entire LIRG and ULIRG populations out to
   redshifts of $\sim0.4$ and $\sim1$, respectively.\label{fig:limits}}}
\end{figure}

\begin{figure}
{\center
\includegraphics[bb =  14 14 335 234, width=\columnwidth] {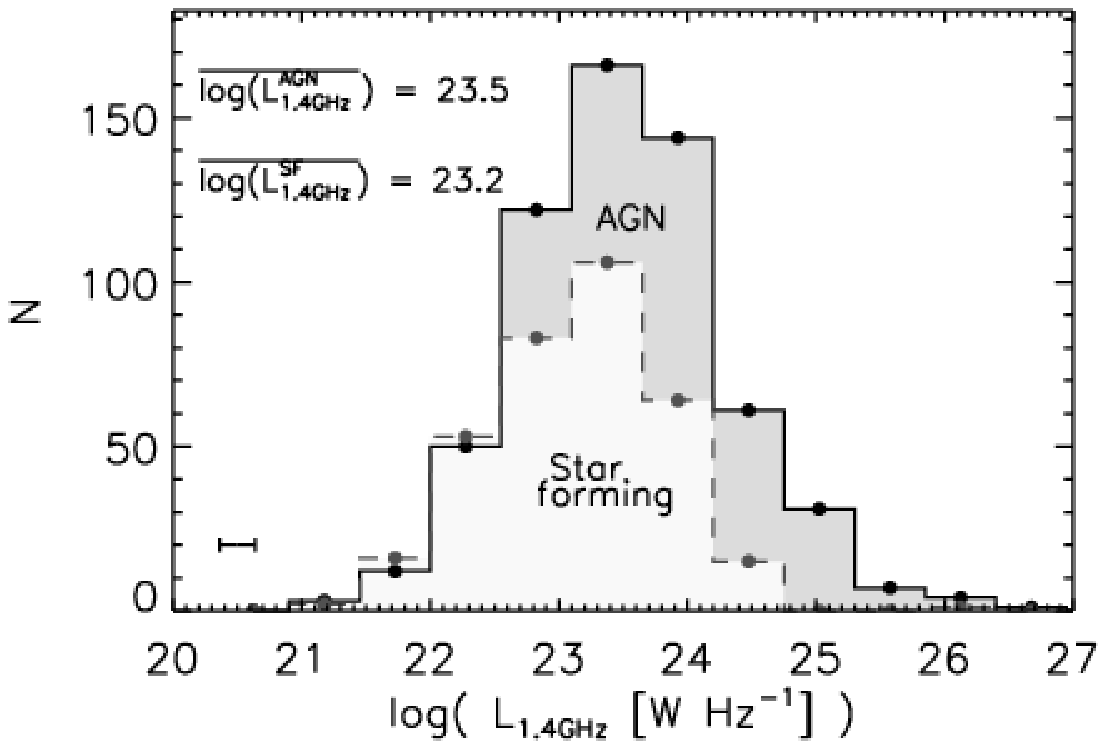}\\
\includegraphics[bb =  14 14 426 196, width=\columnwidth] {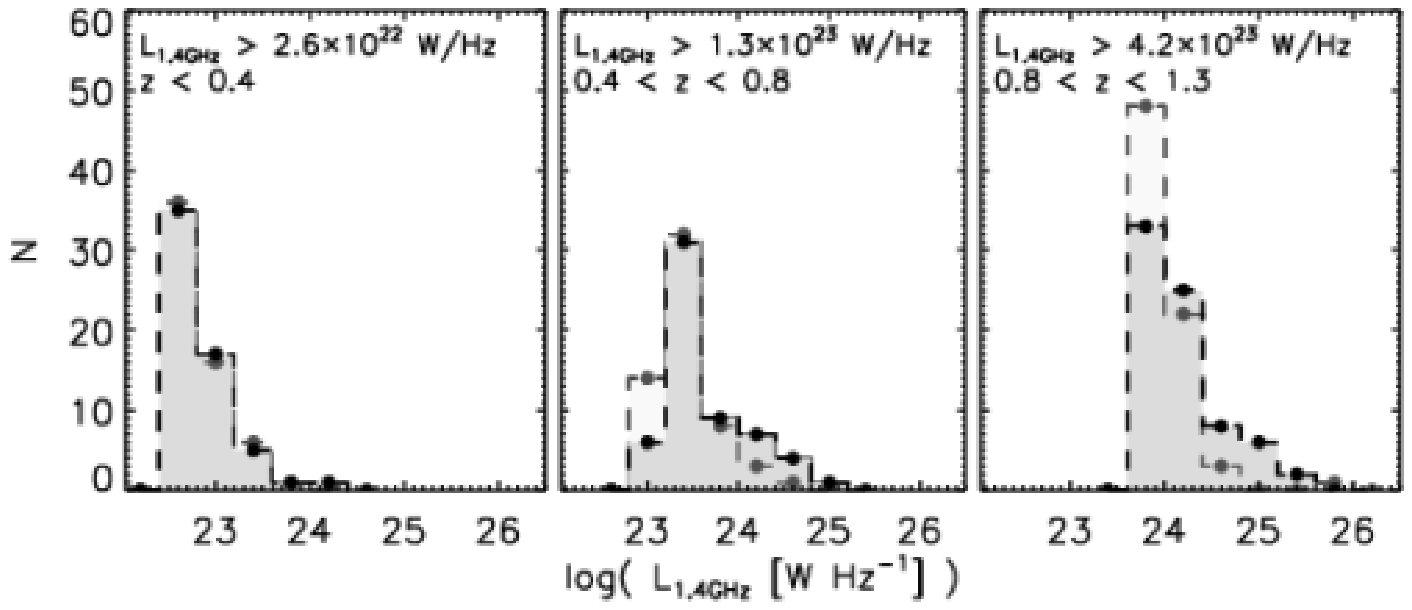}
 \caption{  Distribution of the logarithm of the 1.4~GHz luminosity for the
     selected star forming (light-grey shaded histograms) and AGN (dark-grey
     shaded histograms) galaxies for the entire flux limited sample (top
     panel), and for volume limited samples in 3 redshift ranges (bottom 3
     panels). In the top panel the median $\log{ \left(
         \mathrm{L}_{1.4\mathrm{GHz}} \right) }$ values for the full star
     forming and AGN samples, and the average error in $\log{ \left(
         \mathrm{L}_{1.4\mathrm{GHz}} \right) }$ are indicated. The redshift
     and luminosity ranges of the volume limited samples are given in the
     bottom 3 panels.}
   \label{fig:lsfagn}}
\end{figure}
Further, the luminosity distribution shown in \f{fig:lsfagn} \ agrees well
with local results, which have shown that $\mathrm{L}_{1.4\mathrm{GHz}}$ for
star forming and (absorption and emission line) AGN shows overlapping
distributions, and consequently no clear separation
(e.g.\ \citealt{sadler99, jackson00, chan04}).  
%In general, for our VLA-COSMOS
%SF and AGN galaxies we find no significant difference between the mean values
%of $\mathrm{L}_{1.4\mathrm{GHz}}$, which correspond to
%$\sim2\times10^{23}$~W\,Hz$^{-1}$ and $\sim3\times10^{23}$~W\,Hz$^{-1}$,
%respectively.  
In the local universe \citet{sadler99}
%, based on galaxies observed by the 2dF Galaxy Redshift Survey and the NVSS
%out to $z\sim0.3$, 
inferred a median $\mathrm{L}_{1.4\mathrm{GHz}}$ for SF and AGN galaxies to be
$\sim3\times10^{22}$~W\,Hz$^{-1}$ and $\sim3\times10^{23}$~W\,Hz$^{-1}$,
respectively. Hence, our median value for the luminosity of VLA-COSMOS AGN
($3.2\times10^{23}$~W\,Hz$^{-1}$) out to $z=1.3$ matches the one inferred
locally, however for SF galaxies ($1.6\times10^{23}$~W\,Hz$^{-1}$) it is
higher than that derived by \citet{sadler99}.  The latter is easily understood
as the combined effect of the higher median redshift ($\sim0.7$) of the
galaxies in our {\em flux limited} sample (thus not probing low
$\mathrm{L}_{1.4\mathrm{GHz}}$) and of the higher level of star formation
activity, which is observed going from redshift 0 to 1 \citep{madau96,
  hopkins04} also implying higher $\mathrm{L}_{1.4\mathrm{GHz}}$
\citep{condon92,bell03}.

  We caution at this point that luminosity distributions of flux limited
  samples are strongly dependent on the detection limits, and their
  interpretation has to be approached carefully. For this reason, in
  \f{fig:lsfagn} \ we also show the luminosity distributions for volume
  limited samples of our selected SF and AGN galaxies in three redshift
  ranges. On average AGN galaxies have higher 1.4~GHz luminosities than SF
  galaxies, reassuring the validity of our selection method. Note, however, that
  in the lowest redshift range the AGN and SF galaxy distributions are very
  similar. This is not surprising, but rather consistent with the observed
  luminosity range, that encompasses the region of indistinguishable space
  density of SF and AGN galaxies in the local universe (see local 1.4~GHz
  luminosity functions for SF and AGN galaxies; e.g. \citealt{best04}). 

\subsection{ The 'population mix' in the VLA-COSMOS survey  }
\label{sec:popmix}
In this Section we study the contribution of different sub-populations to the
the total sub-mJy radio population. The key question we want to answer is: Is
the sub-mJy population dominated by any particular sub-population, which may
be the main cause for the observed flattening of the differential radio source
counts below 1~mJy (for VLA-COSMOS source counts; see \citealt{bondi07})?

\subsubsection{ Is the matched radio sample at sub-mJy levels dominated by star
  forming galaxies? } 

In order to obtain an insight into the 'population mix'
of the faint radio sources in the matched radio source sample, in
\f{fig:fluxhisto} \ we show the
%differential (top panel), fractional (middle panel), and cumulative (bottom
%panel) 
distribution of the 1.4~GHz total flux density, $\mathrm{F_{tot}}$, for the SF
and AGN galaxies in the matched radio sample out to $z=1.3$, as well as for
the identified QSOs.\footnote{Photometric redshift information for QSOs in the
  COSMOS survey is not available at this point.}  Note, that the remaining
$z>1.3$ galaxies in the matched radio sample are defined as high redshift
(high-z) galaxies. The flux bins in \f{fig:fluxhisto} \ are 0.15~mJy wide.
Such wide bins allow us to study the average behavior of the galaxies in the
sub-mJy population with decreasing flux densities. Our main aim is to answer one of
the more controversial questions in radio astronomy: {\em Is the sub-mJy radio
  population dominated by star forming galaxies or any other distinct
  sub-population?}

\begin{figure}
{\center
\includegraphics[bb = 14 14 234 406, width=\columnwidth] {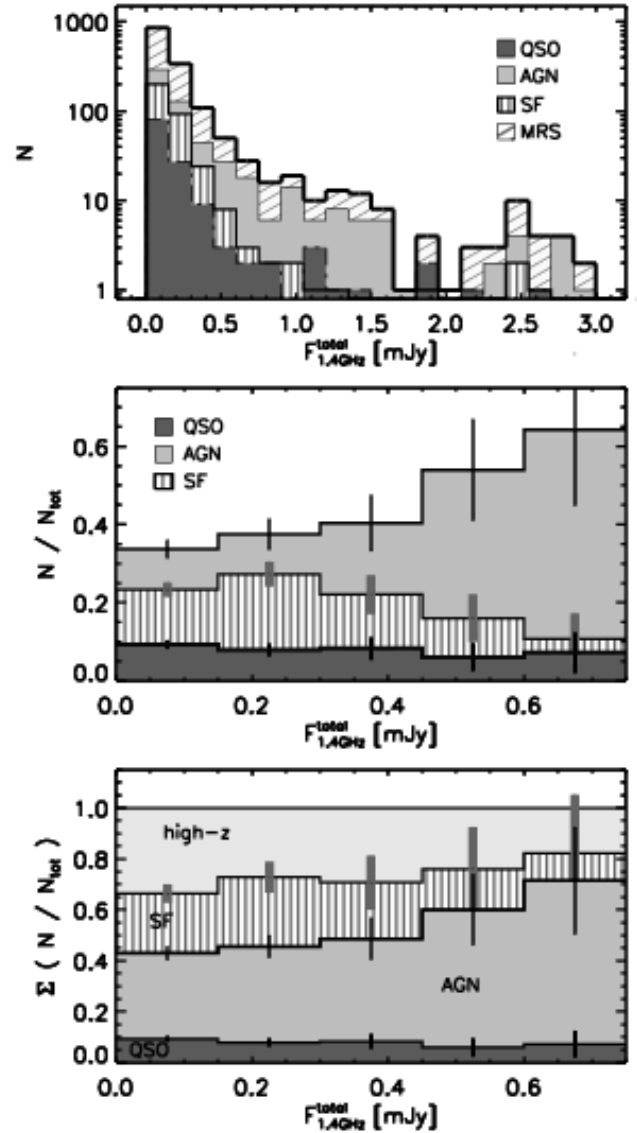}
 \caption{ {\em Top:} The distribution of the 1.4~GHz total flux for the star
   forming (vertically hatched histogram) and AGN galaxies (light-grey shaded
   histogram) in the matched radio sample out to $z=1.3$ (see also
   \f{fig:lsfagn} ). Shown is also the distribution for the selected QSOs
   (dark-grey shaded histogram), for which good redshift information is not
   available at this point.  'MRS' stands for the 'matched radio sample'. {\em
     Middle:} The relative contribution of SF, AGN galaxies out to $z=1.3$,
   and QSOs to the matched radio sample. Note that the missing fraction of
   sources consists of the high redshift (high-z) objects, which we define as
   galaxies in our matched radio sample with $z>1.3$. {\em Bottom:} The
   cumulative distribution of SF, AGN galaxies out to $z=1.3$, QSOs and high-z
   galaxies in the matched radio sample. The indicated error bars in the
   middle and bottom panels are derived from Poisson statistics. Note also the
   different flux scales in the top and the middle/bottom panels.
   \label{fig:fluxhisto}}}
\end{figure}

From the top panel in \f{fig:fluxhisto} \ it is obvious that at flux densities
above $\sim0.7$~mJy we are hampered by low number statistics (the total number
of sources in each bin is below 20).  Therefore, for the purpose of this paper
we will focus only on sources with flux densities below $0.7$~mJy down to the
VLA-COSMOS $5\sigma$ limit of $\sim50~\mathrm{\mu}$Jy. 

The middle panel in \f{fig:fluxhisto} \ shows the fractional distribution of
the identified populations in the matched radio sample, with indicated Poisson
errors. Our findings are as follows. QSOs contribute to the matched radio
sample at a constant level of about $10\%$.  AGN galaxies below $z=1.3$ show a
decreasing trend with decreasing flux densities, with their contribution to the
matched radio sample dropping from $\sim60\%$ to less than $40\%$. The SF
galaxies at $z\le1.3$ in the flux density range of $50~\mu$Jy to $0.7$~mJy contribute
fairly constantly at the given flux densities, with an average contribution of about
$20\%$. Note that the possible increment from $0.7$~mJy to $50~\mathrm{\mu}$Jy
of only $\sim10\%$ is not significant. In \s{sec:testRF} \ we have inferred
that the photometric errors in the synthesized \pone\ color introduce a
positive bias of $\sim5\%$ for SF galaxies. As this bin contains the lowest
number of SF galaxies, it is the most affected by this bias.  In
\s{sec:loccompl} , based on the local SDSS/NVSS sample of galaxies, we have
shown that our \RFmethod\ selects $\sim70\%$ of 'real' SF galaxies, which
make-up $\sim85\%$ of the complete sample of SF galaxies, while the
photometrically selected AGN sample is contaminated by SF galaxies at the
$5\%$ level. Assuming that the percentages of completeness and contamination,
derived from the analysis of the SDSS/NVSS sample of galaxies, can be safely
applied also to our VLA-COSMOS sample, we can use them to correct the observed
fractions. Even with the correction the fractional contribution of SF galaxies
at $z<1.3$ in the matched radio population essentially stays the same, i.e.\ 
about $\sim20\%$ at flux densities in the range from $50~\mathrm{\mu}$Jy to $0.7$~mJy.
Contrary to previous studies (e.g.\ \citealt{seymour04, benn93}), our results
show that SF galaxies at intermediate redshifts are not the dominant
population at sub-mJy flux density levels.  However, it may be possible that a
significant number of SF galaxies at $z>1.3$ exists, which may contribute
strongly to the sub-mJy population. We investigate this possibility in the
bottom panel in \f{fig:fluxhisto} \ and in \f{fig:stern4matched} , and show
below that this is not the case.

\begin{figure}
{\center
\includegraphics[bb = 14 14 338 280, width=\columnwidth]  {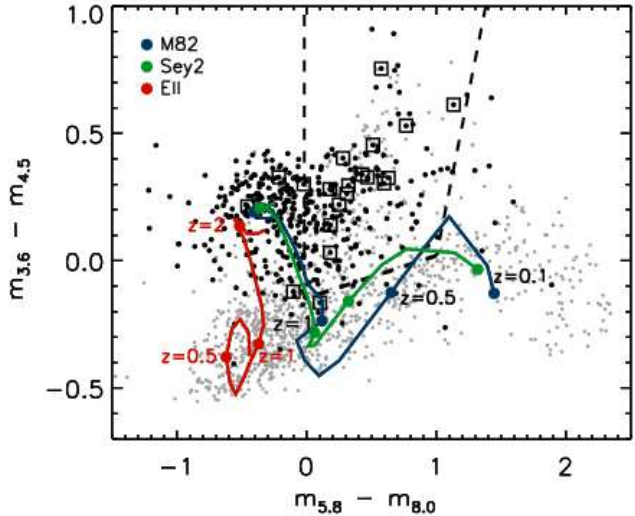}
 \caption{ MIR color -- color digram (analogous to \f{fig:lacystern} \ , bottom
   panel) for the full matched radio sample. Gray dots represent sources
   classified using our \CLASSmethod\ (i.e.\ SF and AGN galaxies out to
   $z=1.3$, and QSOs), while the black dots show the high-z sources (galaxies
   in the matched radio sample beyond $z=1.3$). Open black squares represent
   the high-z sources that have an XMM point source counterpart. The dashed
   lines are indicated to guide the eye, and correspond to the AGN type-1
   selection region, proposed by \citet{stern05}. The curved green, blue, and
   red lines correspond to the color-color tracks, obtained from SEDs of the
   starburst galaxy M~82, a composite of 28 Seyfert~2 galaxies, and a 13~Gyr
   old elliptical galaxy, respectively, in the redshift range from $0.1$ to
   $2.5$.
   \label{fig:stern4matched}}}
\end{figure}

About $30\%$ of the matched radio sample consists of galaxies at redshifts
beyond $z=1.3$ (high-z; see bottom panel in \f{fig:fluxhisto} ).  The
contribution of these galaxies marginally increases from 20\% at 0.7~mJy to
35\% at 50~$\mu$Jy. Although we were not able to apply our \RFmethod\ to these
galaxies in order to classify them, we can nevertheless draw some conclusion
about their nature by studying their multi-wavelength properties. Therefore,
in \f{fig:stern4matched} \ we plot the MIR diagnostic diagram, for the high-z
galaxies and compare their properties to the classified sources in the matched
radio sample.  We overlay non-evolving color tracks for the typical starburst
galaxy M82, an elliptical 13~Gyr old galaxy, and a Seyfert-2 type SED,
obtained from a composite spectrum of 28 randomly chosen Seyfert galaxies
\citep{polletta06}. As expected, the properties of high-z galaxies are
consistent with properties of higher redshift ($z>1.5$) galaxies.  However,
the region they occupy in this MIR color-color diagram is occupied in a
similar way by different sub-populations, such as star forming, Seyfert-type,
and passively evolving galaxies, in the redshift range from about $1.5$ to $3$
(see color tracks in \f{fig:stern4matched} ).  Further, it is possible that a
fraction of these sources are broad line AGN as they are located within the
area outlined by dashed lines, which was proposed by \citet{stern05} for the
selection of AGN Type-1 (see also \citealt{caputi07}). This is further
strengthened by the 22 high-z sources which were also identified as X-ray
point sources, and lie within the Type-1 AGN selection region. Further, given
the high redshift of these sources, combined with the XMM detection limit, it
is highly unlikely that any of these sources may be star forming (see e.g.\ 
Fig.~14 in \citealt{trump07}). This analysis alone already indicates that the
matched radio sources beyond $z=1.3$ are most probably a mixture of different
sub-populations.

The population mix in the high-z sample is further affirmed by the
distribution of the high-z galaxies in the $BzK$ diagram
(shown in \f{fig:bzk4matched} ), which is commonly utilized for the selection
of $z>1.4$ galaxies, and separates well passively-evolving galaxies from those
originally defined as 'star forming' at $z>1.4$ \citep{daddi04}.  Form
\f{fig:bzk4matched} \ it becomes obvious that the BzK criterion does not
select a pure 'star forming' sample at $z>1.4$, as initially postulated, but a
sample that is comprised of both SF galaxies and low-luminosity AGN. In
addition, the X-ray detected high-z sources, which may be classified as AGN
with high confidence as discussed above, also lie within this region.
Therefore, the SED color tracks combined with the distribution of the high-z
galaxies (black dots in \f{fig:bzk4matched} ), and the X-ray detected
subsample suggest that our $z>1.3$ matched sources are a fair mixture of SF
and AGN galaxies, yielding that the high-z galaxies continue to consist of
{\em different} galaxy populations at higher redshifts.

\begin{figure}
{\center
\includegraphics[bb =  14 14 320 271, width=\columnwidth]  {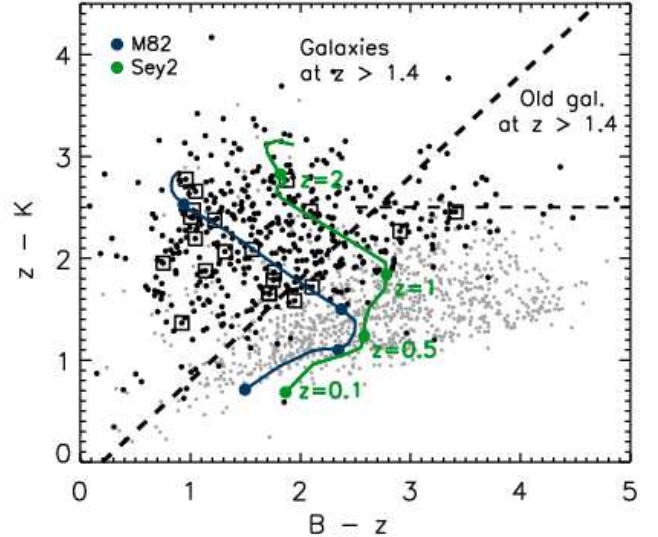}
 \caption{ $B-z$ vs. $z-K$ color-color diagram for the entire matched radio
   sample. The symbols are analogous to \f{fig:stern4matched} : Gray dots
   represent sources identified using our \CLASSmethod\ (i.e.\ SF and AGN
   galaxies out to $z=1.3$, and QSOs), the black dots show the high-z sources,
   and open squares show XMM detected point sources in the high-z sample. The
   dashed lines separate regions of passively evolving (top right; outlined by
   the diagonal and horizontal dashed lines) and star-forming galaxies (left
   of the diagonal dashed line) at $z>1.4$, adopted from \citet{daddi04}.  The
   curved green and blue lines correspond to the color-color tracks, obtained
   from SEDs of the starburst galaxy M~82, and a Seyfert~2 composite,
   respectively, in the redshift range from $0.1$ to $2.5$. Note, that Seyfert
   type of galaxies at $z>1.4$ are also present in the region of 'star-forming
   galaxies', as initially postulated by \citet{daddi04}.
   \label{fig:bzk4matched}}}
\end{figure}

For reference, the $i$ band magnitude distribution for the identified
sub-samples in the matched radio sample is shown in \f{fig:imaghisto} . Note
that the high-z galaxies are the faintest optical sources, consistent with the
expectations for high redshift sources drawn from a flux limited sample.
However, we want to mention that even in the most extreme case that all of the
high-z galaxies were star forming, still the fractional contribution of star
forming galaxies to the matched radio source population would not dominate
over the contribution of AGN and QSOs, but the two contributions would rather
be comparable.

\begin{figure}
{\center
\includegraphics[bb = 14 14 436 381, width=\columnwidth] {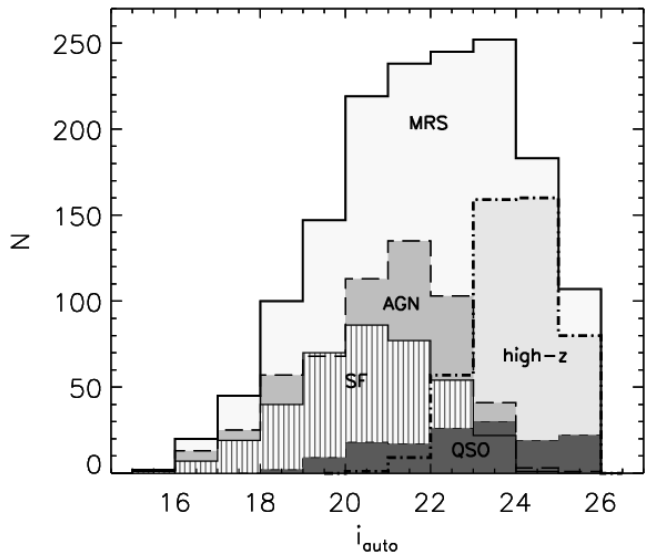} 
\caption{ Distribution of the $i$ band magnitude (Subaru where available,
  otherwise CFHT) for sources in the VLA-COSMOS matched radio sample (MRS).
  Also shown are the distributions for the identified sub-samples of sources:
  QSOs, star-forming (SF), AGN, and high-z galaxies\label{fig:imaghisto}}}
\end{figure}

As shown in \f{fig:bzk4matched} \ it is likely that both classes of objects
contribute similarly to the high-z galaxies, although their exact relative
fractions cannot be determined from this analysis. However, if we assume that
roughly $50\%$ of the high-z galaxies are SF galaxies, then the fraction of SF
galaxies in the matched radio sample would increase to less than $40\%$.
Therefore, we conclude that the population of star forming galaxies is not the
dominant population at sub-mJy levels, at least in our sample of radio sources
with optical counterparts out to $i=26$.  

Our results are strikingly similar to the recent results by
\citet{padovani07}, who studied the sub-mJy radio source population in the
Chandra Deep Field South down to a $5\sigma$ limit of $42~ \mathrm{\mu}$Jy.
They found that SF galaxies make up only about $20\%$ to $45\%$ (i.e.\ roughly
1/3) of the sub-mJy radio sources.

\subsubsection{ The contribution to  the sub-mJy radio population from radio
  sources with no, or with flagged, optical counterparts }
 
Most of the previous studies (as well as our study up to this point) of the
faint radio population have relied on sub-samples of the observed radio
sources, that have been identified with optical counterparts out to $i=26$, as
representative of the entire sub-mJy radio population. For example,
\citet{benn93} used a sample of only 87 out of 523 (i.e.\ less than $20\%$)
1.4~GHz radio sources above 0.1~mJy, for which they obtained optical
spectroscopy ($B<22$), to conclude that above 1~mJy about $50\%$ of the
galaxies were SF or Seyfert galaxies, while below 1~mJy the fraction increases
to $\sim90\%$.  Further, \citet{gruppioni99} studied optical spectroscopic
properties of 34 radio sources above $0.2$~mJy in the Marano Field down to
$B=24$. This sample comprised $\sim60\%$ of the entire sample of faint radio
sources, and they concluded that the SF galaxies do not constitute the main
galaxy population of their radio sources, and even at sub-mJy levels the
majority of their radio sources were identified with early type galaxies,
consistent with AGN.  Gruppioni et al.\ attributed the difference in their
results compared to the results from \citet{benn93} to the fainter optical
magnitude limit reached for their radio sample. In this work we have an order
of magnitude larger sample size (\no\ sources with optical counterparts), and
a significantly deeper optical limit ($i=26$) than previous studies.  However,
still our matched radio source sample consists of only $\sim65\%$ of the
VLA-COSMOS 1.4~GHz sources.  Therefore, it is important to investigate the
contribution of the remaining $\sim35\%$ of the radio sources, with no
identified or flagged optical counterparts, to the sub-mJy population. It may
indeed be possible that a 'missing' population of objects, that significantly
contributes to the sub-mJy population, is 'hidden' in this sample. We show
below that this is not the case.
 
If the above hypothesis is true, then the properties (such as e.g.\ MIR
colors) of these remaining objects are expected to be distinct from the
properties of objects in the matched radio sample.  In order to shed light on
this, in \f{fig:fluxhisto4rem} \ (top panel) we show the distribution of the
total flux density for the \no\ sources in the matched radio sample, and for the
remaining \noREM\ sources that were a) not identified with optical
counterparts with $i\leq26$ or b) have optical counterparts with $i\leq26$ but
in photometrically masked-out region (see \s{sec:CCradio-opt} ).  The bottom
panel shows the fractional contribution of these two samples compared to the
entire VLA-COSMOS 1.4~GHz population.  The fraction of matched radio sources
is statistically consistent to be constant at $\sim65\%$ at all faint flux density
levels, although formally it decreased from $\sim75\%$ at $\sim0.7$~mJy to
$\sim60\%$ at the limit of the VLA-COSMOS survey. On the basis of these high
percentages, it is unlikely that any further population, that is not present
in our matched radio sample, could account for a significant fraction of the
sub-mJy population.
%
% 22
\begin{figure}
{\center
\includegraphics[width=\columnwidth]  {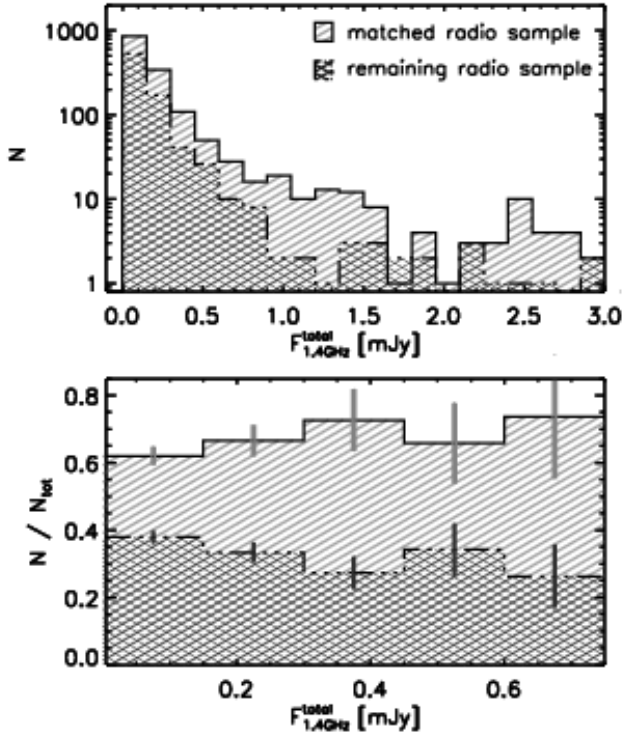}
 \caption{ {\em Top:} The distribution of the 1.4~GHz total flux for sources
   in the matched radio sample (single hatched histogram), and for other
   VLA-COSMOS sources that have either no identified, or a photometrically
   flagged, optical counterpart brighter than $i=26$ (cross-hatched
   histogram). {\em Bottom:} The fractional distribution of the two samples
   compared to the entire sample of 1.4~GHz radio sources. Indicated error
   bars are derived from Poisson statistics.
   \label{fig:fluxhisto4rem}}}
\end{figure}
\begin{figure}
{\center
\includegraphics[width=\columnwidth]  {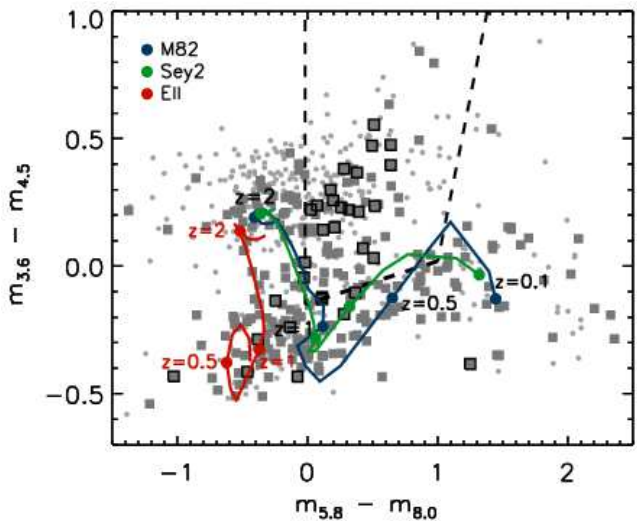}
 \caption{ Analogous to \f{fig:stern4matched} , but for $813$ VLA-COSMOS
   1.4~GHz sources (grey dots) that have either no identified, or a
   photometrically flagged, optical counterpart brighter than $i=26$, but that
   have IRAC detections (i.e.\ $\sim75\%$; see \s{sec:CC_no-counterparts} ).
   Grey squares indicate the sources that have an optical counterpart with
   $i\leq26$, but that is within a masked-out area. Black open squares
   represent the sources that have XMM point source detections.  The clumping
   of objects in the upper left quadrant is consistent with both star forming,
   Seyfert and passive galaxies, in the redshift range from about $1.5$ to
   $3$, suggesting that over $\sim60\%$ of the VLA-COSMOS 'remaining' sources
   are high redshift galaxies or QSOs.
   \label{fig:sternRem}}}
\end{figure}

This is further strengthened by the distribution of the remaining sources in
the MIR color-dolor diagram, shown in \f{fig:sternRem} , which is consistent
with the expected distribution for a mixed sample of star forming, AGN
galaxies, and QSOs at all redshifts in the range from the local to the highest
observable redshifts. An additional affirmation of the mix of population
arises from the 31 sources that have been detected as X-ray point sources (see
\f{fig:sternRem} ). It is worth noting however, that $\sim60\%$ of these
sources are consistent with higher redshift objects ($z\gtrsim1.3$), while
this is true only for $\sim50\%$ of the galaxies in the matched radio sample
(see \f{fig:stern4matched} ). We therefore conclude that the radio sources
without identified, or with flagged, optical counterparts brighter than $i=26$
are most likely comprised of a mixture of different source populations (SF,
AGN, QSO), similar to, although on average at higher redshift then the radio
sources in our matched radio sample.  Further, the relatively small total
percentage of these sources cannot significantly alter the results about the
'population mix' in the 1.4~GHz VLA-COSMOS radio sample, inferred in the
previous section.

%\comm{ Summarize the limits of the classification method: faint radio levels,
%  optical/redshift limits. }

\subsection{ Concluding remarks on the composition of the faint radio
  population  } 
\label{sec:discussion}

The faint -- submillijansky -- radio population comprises the radio population
responsible for the upturn of the differential radio source counts below 1~mJy
(see e.g.\ \citealt{bondi07}), and it has been the subject of many studies,
and a matter of great debate in the past three decades \citep{condon84,
  windhorst85, gruppioni99, seymour04, simpson06}. This radio population has
been interpreted as a 'new' rising population of objects, that do not
significantly contribute at higher radio flux density levels. However, results from
studies that tried to reveal the exact composition of the sub-mJy population
have been highly discrepant. It was suggested that the majority of this
faint radio population consists of faint blue galaxies, and it was assumed
that these galaxies are undergoing significant star formation
\citep{windhorst85}. The spectroscopic study by \citet{benn93}, although
analyzing only less than $20\%$ of their radio sample, supported this
result indicating that the fraction of SF and Seyfert galaxies rises from
about $50\%$ to $90\%$ in the range from super- to sub-mJy flux densities. However,
\citet{gruppioni99}, who performed a deeper optical spectroscopic analysis of
a larger fraction ($\sim60\%$) of their radio sources, disagreed with this
result identifying the majority of their sub-mJy radio sources with early type
galaxies, in which the radio emission is produced by AGN. Recently, using a
combination of optical and radio morphology as an identifier for AGN and SF
galaxies on $\sim90\%$ of their radio sources in the SSA13 field,
\citet{fomalont06} suggested that less than $\sim40\%$ of the radio sources
below 1~mJy are AGN. On the other hand, \citet{padovani07} based their SF/AGN
classification on a combination of optical morphologies, X-ray luminosities
and radio--to--optical flux ratios of their radio sources in the CDFS (Chandra
Deep Field South), and indicated that only about $20-40\%$ of the faint radio
sources are made-up of star forming galaxies.

As already mentioned in the \s{sec:intro} \ there are two main reasons for
such controversial results in the past literature: a) The identification
fraction of radio sources with optical counterparts spans a wide range (20\%
to 90\%) in samples from radio deep fields, and b) the methods for the
separation of SF from AGN galaxies have been highly heterogeneous.  Further,
the first and second point seem to exclude each other.  Namely, the most
efficient separation methods (e.g.\ intermediate- to high- resolution optical
spectroscopy) generally led to a small identification fraction, whereas large
radio samples with high optical identification fractions usually lacked robust
SF/AGN indicators. In this paper we have tried to make a strong conjunction of
these two points. We introduced a new method to separate SF from AGN galaxies,
which relies strongly on rest-frame color properties, and thus allowed us to
perform a robust classification of $\sim65\%$ ($=\no$) of the 1.4~GHz
VLA-COSMOS radio sources down to $\sim50~\mathrm{\mu Jy}$ with identified
optical counterparts out to $i=26$ without the need for e.g.\ optical
spectroscopy. Thus, we managed to reach both a high identification fraction
and a robust source classification.  However, in order to avoid any possible
biases, we also utilized the full panchromatic COSMOS data set to put
constraints on the properties of the remaining $\sim35\%$ of the radio sources
that were not identified with optical counterparts with $i\leq26$, or that
have optical counterparts with $i\leq26$ but with uncertain photometry due to
blending and saturation in the optical images.  In short, we have obtained a
complete view of the 'population mix' of the sub-mJy radio sources, using to
date the largest sample of \noSN\ faint ($5\sigma\approx50~\mathrm{\mu}$Jy)
radio sources at 1.4~GHz. Our combined methods allow us to reveal the true
nature of the sub-mJy radio sources in the VLA-COSMOS survey on a sound
statistical basis.

\citet{jarvis04} first suggested that the observed flattening of the
differential radio source counts below 1~mJy may be caused by 'radio-quiet'
AGN, i.e.\ radio-quiet QSOs and type-2 AGN (consistent with our 'AGN' class),
rather than star forming galaxies. Making use of the observed correlation
between X-ray and radio luminosity and the empirically derived X-ray
luminosity function, they computed the 1.4~GHz luminosity function for
radio-quiet AGN out to high redshifts, and used it as the key ingredient to
model the radio source counts at faint flux density levels (see \citealt{jarvis04} and
references therein).  However, they did not take into account SF galaxies as
an important population to explain the shape of the counts. Such an
interpretation managed to represent the differential radio source counts well
down to $\sim250~\mathrm{\mu Jy}$, but failed at fainter flux densities.  Our results,
derived in the previous sections, show that indeed star forming galaxies are
not the dominant population at sub-mJy levels in the range of
$50~\mathrm{\mu}$Jy to $0.7$~Jy, however they still contribute about $30\%$ to
$40\%$ at these flux densities.  Based on our large sample of \noSN\ VLA-COSMOS radio
sources we find that our classified AGN, which comprise mostly of
low-luminosity and type-2 AGN, make up $50-60\%$ of the faint radio population
with a decreasing trend towards fainter flux densities, while the identified QSOs,
which are mostly type-1 AGN, form a minor contribution of $\sim10\%$ of the
sub-mJy radio population in the range of $50~\mathrm{\mu}$Jy to $0.7$~Jy.
Thus, our observational results show that the 'population mix' in the faint
radio population contains a fair contribution of both SF galaxies and
(low-luminosity and obscured) AGN, at least down to $50~\mathrm{\mu}$Jy.
Thus, separate luminosity functions for both populations have to be taken into
account in order to fully explain the flattening of the radio source counts
below 1~mJy.  The 1.4~GHz luminosity function for star forming and AGN
galaxies at intermediate redshift, as well as the modeling of the 1.4~GHz
radio source counts using these new observational constrains and further
analyzes of the 'population mix' in the VLA-COSMOS survey, is going to be
fully addressed and presented in a number of up-coming publications
\citep{smo07, ciliegi07b, paglione07}.

%\subsection { Notes of caution }  
%\label{sec:caution}
% It is worth discussing briefly some of the limitations and caveats of the
% \RFmethod\ presented here. 

\section{ Summary }
\label{sec:summary}

Our newly developed \RFmethod, in conjunction with the VLA-COSMOS 1.4~GHz
  observations, enabled us to make, for the first time, a thorough distinction
  between sources where the 1.4~GHz radio emission is predominantly driven by
  star formation processes from those where it is driven by SMBH processes,
  {\em regardless of the luminosity of the latter}, and apply it to currently
  the largest sample of \no\ 1.4~GHz sources, with optical counterparts
  brighter than $i_\mathrm{AB}\leq26$, and complete down to
  $\sim50~\mathrm{\mu Jy}$. 
  
  We have explored to full detail the composition of the sub-mJy radio
  population making use of the entire sample of \noSN\ VLA-COSMOS radio
  sources detected above $5\sigma~(\approx 50\mathrm{\mu Jy})$, in conjunction
  with the panchromatic (X-ray to radio) COSMOS data set. We find that SF
  galaxies are not the dominant population at submillijansky flux density
  levels, as believed previously, but that they rather make up an
  approximately constant fraction of $30-40\%$ in the flux density range of
  $\sim50~\mathrm{\mu}$Jy to $0.7$~mJy.  The radio population at these flux
  densities is a mixture of roughly $30-40\%$ of SF and $50-60\%$ of AGN
  galaxies, with a minor contribution ($\sim10\%$) of QSOs.

\acknowledgments 

The authors would like to thank M.~Polletta for the spectral template
library.  VS would like to thank E.~Bell, G.~Fabbiano, G.~Helou and
D.~Frayer for insightful discussions, as well as M. Schartmann for his
help regarding Fortan compilers.  CC, ES and VS acknowledge support
from NASA grant HST-GO-09822.31-A.  CC would like to acknowledge
support from the Max-Planck Society and the Alexander von Humboldt
Foundation through the Max-Planck-Forschungspreis 2005.  KJ
acknowledges support by the German DFG under grant SCHI 536/3-1.  CJW
is supported by the MAGPOP Marie Curie EU Research and Training
Network.

This work is based on observations with the National Radio Astronomy
Observatory which is a facility of the National Science Foundation operated
under cooperative agreement by Associated Universities, Inc; the NASA/ESA {\em
  Hubble Space Telescope}, obtained at the Space Telescope Science Institute,
which is operated by AURA Inc, under NASA contract NAS 5-26555; also based on
observations obtained with XMM-Newton, an ESA science mission with instruments
and contributions directly funded by ESA Member States and NASA; also based on
data collected at: the Subaru Telescope, which is operated by the National
Astronomical Observatory of Japan; the European Southern Observatory, Chile;
Kitt Peak National Observatory, Cerro Tololo Inter-American Observatory, and
the National Optical Astronomy Observatory, which are operated by the
Association of Universities for Research in Astronomy, Inc.\ (AURA) under
cooperative agreement with the National Science Foundation. Based also on
observations obtained with MegaPrime/MegaCam, a joint project of CFHT and
CEA/DAPNIA, at the Canada-France-Hawaii Telescope (CFHT) which is operated by
the National Research Council (NRC) of Canada, the Institut National des
Science de l'Univers of the Centre National de la Recherche Scientifique
(CNRS) of France, and the University of Hawaii. This work is based in part on
data products produced at TERAPIX and the Canadian Astronomy Data Centre.

    Funding for the Sloan Digital Sky Survey (SDSS) and SDSS-II has been
    provided by the Alfred P. Sloan Foundation, the Participating
    Institutions, the National Science Foundation, the U.S. Department of
    Energy, the National Aeronautics and Space Administration, the Japanese
    Monbukagakusho, and the Max Planck Society, and the Higher Education
    Funding Council for England. The SDSS Web site is http://www.sdss.org/. 

    The SDSS is managed by the Astrophysical Research Consortium (ARC) for the
    Participating Institutions. The Participating Institutions are the
    American Museum of Natural History, Astrophysical Institute Potsdam,
    University of Basel, University of Cambridge, Case Western Reserve
    University, The University of Chicago, Drexel University, Fermilab, the
    Institute for Advanced Study, the Japan Participation Group, The Johns
    Hopkins University, the Joint Institute for Nuclear Astrophysics, the
    Kavli Institute for Particle Astrophysics and Cosmology, the Korean
    Scientist Group, the Chinese Academy of Sciences (LAMOST), Los Alamos
    National Laboratory, the Max-Planck-Institute for Astronomy (MPIA), the
    Max-Planck-Institute for Astrophysics (MPA), New Mexico State University,
    Ohio State University, University of Pittsburgh, University of Portsmouth,
    Princeton University, the United States Naval Observatory, and the
    University of Washington.

{}

\clearpage
\begin{deluxetable}{cccccc}
\tablecaption{ Multi-wavelength cross-correlation of VLA-COSMOS 1.4~GHz radio sources
     \label{tab:class}}
%\tablewidth{0pt}
   \tablewidth{17cm} 
\tablehead{ 
\colhead{ } & 
\colhead{ total } & 
\colhead{ $z_\mathrm{spec}$ } & 
\colhead{ IRAC\tablenotemark{1} } & 
\colhead{ MIPS\tablenotemark{2} } & 
\colhead{ XMM\tablenotemark{3} } 
}
   \startdata
   total radio sample & 2388 & 520  & 2058 & 1117 & 210 \\
   \hline\hline
   matched radio sample\tablenotemark{*} & 1558 & 447 &  1448 & 799 & 179 \\
   \hline
   stars & 2 & 0 &  2 & 2 & 0 \\
   QSOs & 139 & 31 &  122 & 78 & 43 \\
   SF galaxies\tablenotemark{i}  & 340 & 150 &  322 &  280 & 16  \\
   AGN galaxies\tablenotemark{i} &  601 & 262 & 579 & 267 &  98  \\
   high-z galaxies\tablenotemark{ii} & 476 & 4 &  423 & 172 & 22 \\
   \hline\hline
   remaining radio sample\tablenotemark{**} & 830 & 73 & 610 & 318 & 31
\enddata 
\tablenotetext{1}{ Detected in
     the Spitzer/IRAC 3.6~$\mathrm{\mu}$m band. }
\tablenotetext{2}{
     Detected in the Spitzer/MIPS 24~$\mathrm{\mu}$m band within the shallow
     MIPS COSMOS survey at or above a signal to noise of 3. }
\tablenotetext{3}{ X-ray point-sources associated with optical
     counterparts as described in \citet{brusa07} } 
\tablenotetext{*}{ Radio source sample positionally matched to optical sources
  with AB $i\leq26$ and outside masked-out regions in the NUV-NIR photometric
  catalog described by \citet{capak07}. } 
\tablenotetext{i}{ Galaxies at redshift $\le1.3$. }  
\tablenotetext{ii}{ Galaxies at redshift $>1.3$. }  \tablenotetext{**}{ Radio
  sources a) without optical counterparts with $i\leq26$ ($\sim70\%$), or b)
  with optical counterparts ($i\leq26$) that lie in masked-out areas in the
  photometric catalog ($\sim30\%$). } 
%\tablecomments{ Numbers are still tentative! }
\end{deluxetable}

\begin{appendix}

\section{ The rest-frame colors \pone\ and \ptwo }
\label{app:p1p2}

The rest-frame colors \pone\ and \ptwo\ are a a linear superposition of colors
in the wavelength range of 3500 -- 5800~\AA, obtained using the modified
Str\"{o}mgren photometric system \citep[the filter response curves are
available at http://www.mpia-hd.mpg.de/COSMOS]{odell02,smo06}.  \pone, \ptwo\ 
are principal component axes optimally characterizing the galaxy locus in
2D color-color space.  \pone\ measures the position along the galaxy locus, and
\ptwo\ the position perpendicular to it (see Fig.~4 in \citealt{smo06}).  As
the galaxy locus is slightly curved, the functional form of the rest-frame
colors is given separately for the blue and red ends, with a boundary at
$vz-yz=0.646$.  Hence, for galaxies with $vz-yz\leq0.646$ (\pone,\ptwo) are
given as: \eq{
  \label{Paxes1} P1 = \phantom{X}0.911\,(c_1-0.646) + 0.412\,(c_2-0.261) }
\eq{ P2 = -0.412\,(c_1-0.646) + 0.911\,(c_2-0.261), } and for galaxies with
$vz-yz>0.646$ as: \eq{ P1 = \phantom{X}0.952\,(c_1-0.646) + 0.307\,(c_2-0.261)
} \eq{
\label{Paxes2}
   P2 = -0.307\,(c_1-0.646) + 0.952\,(c_2-0.261),
}

where $c_1=vz-yz$ and $c_2=bz-yz$.  

\section{Derivation of completeness and contamination of the photometrically
  selected samples of star forming galaxies and AGN}
\label{app:complmain}

\subsection{SDSS/NVSS galaxy sample}
\label{app:complfull}

Given both the available spectroscopy and photometry of the local galaxy
sample (SDSS/NVSS) we can easily access the completeness and contamination of
the photometrically selected samples of SF and AGN galaxies as follows. In
\f{fig:p1sdssnvss} \ we show the differential and cumulative distributions of
\pone\ for the SDSS/NVSS star forming, composite and AGN galaxies. Here the
classification is based on the BPT diagram for the galaxies with emission lines
in their spectra, while all galaxies without emission lines are included in
the AGN class. These 'absorption line AGN' constitute the major fraction of
all the AGN ($\sim80\%$) in the sample. The top panel shows the \pone\ 
histograms for these three types of objects, normalized by the total number of
objects.  In the middle panels in \f{fig:p1sdssnvss} \ we show the fractions
for the SF, AGN and composite galaxies for the selection of SF (left panel)
and AGN (right panel) galaxies. The fractions were computed in such a way that
for each \pone\ value ($P1_i$) the distributions were normalized to the total
number of galaxies in the sample with $P1\leq P1_i$ (for SF galaxies; left
panel), or $P1> P1_i$ (for AGN galaxies; right panel). In this way, we obtain
the fraction of SF, AGN and composite galaxies within the full sample that was
selected only by applying a \pone\ color-cut.  Further, in the bottom panels
in \f{fig:p1sdssnvss} \ we show the cumulative distributions of the SF (left
panel) and AGN (right panel) galaxies, scaled to the total number of SF and
AGN galaxies, respectively, thus showing their completeness as a function of
\pone.  Selecting galaxies with \pone$\leq$\ponecut, and defining them as the
'photometrically selected sample of star forming galaxies', leads to a sample
that contains $\sim20\%$ AGN, $\sim10\%$ composite objects, and $\sim70\%$
'real' (i.e.\ spectroscopically identified) SF galaxies (see middle left panel
in \f{fig:p1sdssnvss} ). The latter make-up $\sim85\%$ of {\em all} 'real' SF
galaxies (see bottom left panel in \f{fig:p1sdssnvss} ).  On the other hand,
using the color cut to generate the 'photometrically selected AGN galaxy
sample' (\pone$>$\ponecut) leads to a sample that contains $\sim5\%$ SF
galaxies, $\sim15\%$ composite objects, and $\sim80\%$ 'real' AGN galaxies
(see middle right panel in \f{fig:p1sdssnvss} ), where the latter make-up
$\sim90\%$ of {\em all} 'real' AGN galaxies (see bottom right panel in
\f{fig:p1sdssnvss} ).

\begin{figure}
{\center
\includegraphics{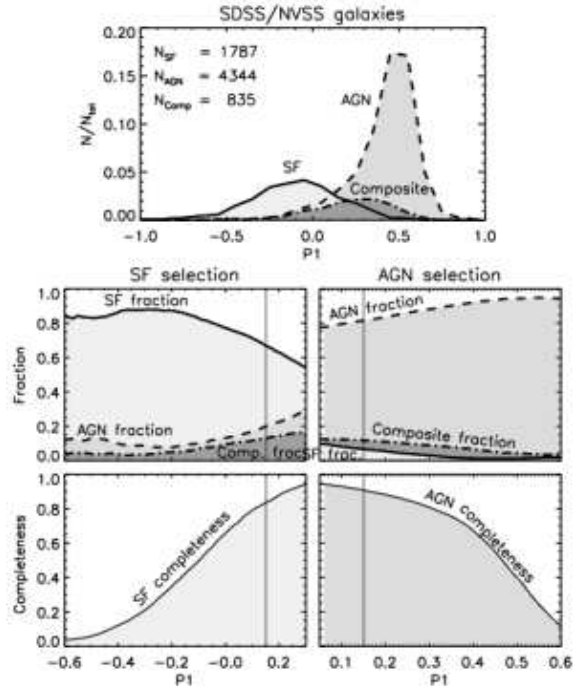}
\caption{ Differential and cumulative distributions of the rest-frame color
  \pone\ for $\sim7,000$ SDSS galaxies, in the redshift range of $0.01$ to
  $0.3$, drawn from the DR1 ``main'' spectroscopic sample, that have 1.4~GHz
  NVSS detections. The galaxies were spectroscopically classified as star
  forming (SF; thick line), AGN (dashed line) or composite (dot-dashed line;
  see text for details). The {\em top panel} shows the \pone\ histograms for
  these three types of objects.  The {\em middle left panel} shows the
  fraction of these three types of galaxies as a function of \pone.
  These distributions are normalized in such a way that for each \pone\ value
  the fraction of the three types of galaxies within the total selected sample
  can be read off. The {\em bottom left panel} shows the cumulative
  distribution of \pone\ for SF galaxies, thus showing to which completeness
  the 'real' (i.e.\ spectroscopically classified) SF galaxies are selected for
  any given \pone\ color-cut.  The {\em bottom and middle right panels} are
  analogous to the bottom and middle left panels, but for AGN galaxies. Note
  also that the cumulative distributions for AGN were computed as a function
  of decreasing \pone. The vertical (thin solid) lines in the middle and
  bottom panels designate the value of \pone, chosen to separate SF galaxies
  from AGN.   \label{fig:p1sdssnvss}}}
\end{figure}

\subsection{SDSS/NVSS/IRAS galaxy sample}
\label{app:compl}
The BPT diagram for $\sim830$ SDSS/NVSS/IRAS emission line galaxies is shown
in \f{fig:bptsdssnvssiras} , color-coded using the \pp\ plane. A strong
correlation between \pone\ and \na , very similar to the correlation
observed for the full sample, is discernible.

The differential distribution of \pone\ for the complete sample of 875
SDSS/NVSS/IRAS galaxies is shown in the top panel of \f{fig:p1sdssnvss} .
The galaxies were spectroscopically separated into AGN, star-forming and
composite galaxies (as described in \s{sec:locsample} ).  In the middle panel
we show the corresponding fraction (analogous to the middle left panel in
\f{fig:p1sdssnvss} ).  Using a rest-frame \pone\ color cut-off of \ponecut\ the
selected sample of star forming galaxies, detected in the IR regime, contains
$\sim70\%$ 'real' (i.e.\ spectroscopically identified) SF galaxies (that make
up $\sim85\%$ of {\em all} 'real' SF galaxies; see bottom panel), $10\%$ AGN,
and $20\%$ composite objects. This is fairly consistent with the properties of the
entire SDSS/NVSS sample.

In order to assess the fraction of dusty starburst galaxies that we omit using
the rest-frame color selection, in the bottom panel of \f{fig:p1sdssnvssiras}
\ we show the cumulative distributions of the spectroscopically identified
star forming galaxies as a function of \pone\ for i) all star forming
galaxies, ii) luminous IR galaxies (LIRGs; $\mathrm{L_{IR}}=10^{11-12}$~\lsun)
and iii) ultra-luminous IR galaxies (ULIRGs, $\mathrm{L_{IR}}>10^{12}$~\lsun).
[The total IR luminosities were computed following \citet{sanders96}.] A
\pone\ cut of \ponecut\ misses only $\sim10\%$ and $\sim5\%$ of luminous and
ultra-luminous starburst galaxies, respectively.

\begin{figure}
{\center
  \includegraphics{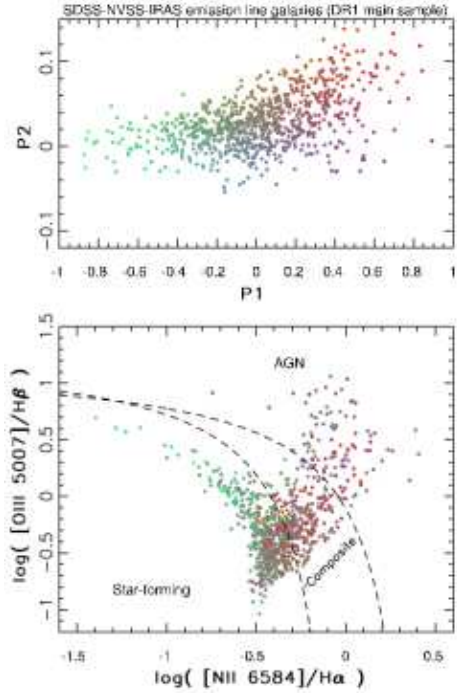}
\caption{ Analogous to \f{fig:bptsdssnvss} \ but for $\sim830$ SDSS/NVSS/IRAS
  emission line galaxies.
\label{fig:bptsdssnvssiras}}}
\end{figure}
\begin{figure}
{\center
  \includegraphics{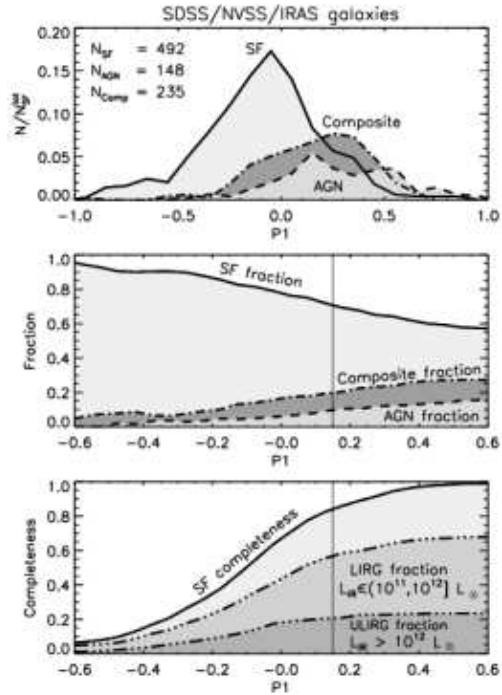}
\caption{ Differential and cumulative distributions of the rest-frame color
  \pone\ for $875$ SDSS galaxies from the DR1 ``main'' spectroscopic sample,
  that have NVSS and IRAS detections. The galaxies were spectroscopically
  classified as star forming (SF; thick line), AGN (dashed line) or composite
  (dot-dashed line; see text for details), and their \pone\ histograms are
  shown in the {\em top panel}. The {\em middle panel} shows the 
  fraction of \pone\ for SF, AGN and composite galaxies, analogous to the
  middle left panel in \f{fig:p1sdssnvss} . The vertical thin solid line shows
  the chosen \pone\ color cut-off. The {\em bottom panel} shows the cumulative
  distribution of \pone, normalized to SF galaxy counts (thick line). Also
  shown are the \pone\ distributions of SF galaxies with IR luminosities
  greater than ~$10^{11}$~\lsun\ and $10^{12}$~\lsun\, consistent with LIRGs
  and ULIRGs, respectively. Note that our \RFmethod\ is not strongly biased
  against dusty starburst galaxies.
\label{fig:p1sdssnvssiras}}}
\end{figure}

\end{appendix}

\end{document}